\documentclass[11pt]{article}
\pdfoutput=1
\newcommand{\Comment}[1]{{}}
\usepackage{amsfonts,amsthm,amsmath,amssymb,slashed}
\usepackage[textwidth = 430 pt, textheight = 630 pt]{geometry}
\usepackage{color}
\usepackage{booktabs}
\usepackage{subcaption}
\usepackage[export]{adjustbox}

\Comment{\usepackage{color}
\definecolor{MyDarkBlue}{rgb}{0.15,0.15,0.45}
\usepackage[linktocpage=true]{hyperref}
\hypersetup{
colorlinks=true,
citecolor=MyDarkBlue,
linkcolor=MyDarkBlue,
urlcolor=MyDarkBlue,
pdfauthor={Luis Alejo and Horatiu Nastase},
pdftitle={Particle-vortex duality and theta terms in AdS/CMT applications},
pdfsubject={hep-th}
}
\usepackage[utf8]{inputenc}
\usepackage[numbers,sort&compress]{natbib}
\usepackage{hypernat}}
\usepackage{graphicx}
\usepackage{cite}
\usepackage[colorlinks=true]{hyperref}
\hypersetup{
  allcolors = blue,
}

\newcommand\ignore[1]{}
\def\one{{\,\hbox{1\kern-.8mm l}}}

\def\a{\alpha}\def\b{\beta}

\def\d{\partial}

\def\R{\mathbb{R}}

\newcommand{\Cset}{{\,\,{{{^{_{\pmb{\mid}}}}\kern-.45em{\mathrm C}}}}}

\newcommand{\be}{\begin{equation}}
\newcommand{\bea}{\begin{eqnarray}}

\newcommand{\ee}{\end{equation}}
\newcommand{\eea}{\end{eqnarray}}

\begin{document}

\renewcommand{\thefootnote}{\fnsymbol{footnote}}

\makeatletter
\@addtoreset{equation}{section}
\makeatother
\renewcommand{\theequation}{\thesection.\arabic{equation}}

\rightline{}
\rightline{}




\begin{center}
{\LARGE \bf{\sc S-duality, entropy function and transport in $AdS_4/CMT_3$}}
\end{center}
 \vspace{1truecm}
\thispagestyle{empty} \centerline{
{\large \bf {\sc Luis Alejo${}^{a},$}}\footnote{E-mail address: \Comment{\href{mailto:luis.alejo@unesp.br}}{\tt luis.alejo@unesp.br}}
{\large \bf {\sc Prieslei Goulart${}^{b}$}}\footnote{E-mail address: \Comment{\href{mailto:prieslei@if.usp.br}}{\tt prieslei@if.usp.br}}
{\bf{\sc and}}
{\large \bf {\sc Horatiu Nastase${}^{a}$}}\footnote{E-mail address: \Comment{\href{mailto:horatiu.nastase@unesp.br}}{\tt horatiu.nastase@unesp.br}}
                                                        }

\vspace{.5cm}


\centerline{{\it ${}^a$Instituto de F\'{i}sica Te\'{o}rica, UNESP-Universidade Estadual Paulista}}
\centerline{{\it R. Dr. Bento T. Ferraz 271, Bl. II, Sao Paulo 01140-070, SP, Brazil}}
\vspace{.3cm}
\centerline{{\it ${}^b$Instituto de F\'{i}sica, Universidade de S\~ao Paulo,}}
\centerline{{\it Rua do Mat\~ao, 1371, 05508-090 S\~ao Paulo, SP, Brazil}}

\vspace{1truecm}

\thispagestyle{empty}

\centerline{\sc Abstract}

\vspace{.4truecm}

\begin{center}
\begin{minipage}[c]{380pt}
{\noindent In this paper we consider Abelian vector plus scalar holographic gravity models for 2+1 dimensional condensed 
matter transport, and the effect of S-duality on them.  We find the transport coefficients from the electric and heat currents
via usual membrane paradigm-type calculations, 
and the effect of S-duality on them. We study the same system also by using the entropy function formalism in the extremal
case, and the formalism 
of holographic Stokes equations, in the case of one-dimensional lattices. We study a few generalizations that appear when 
considering a supergravity-inspired model, and apply the entropy function method for them.
}
\end{minipage}
\end{center}

\vspace{.5cm}

\setcounter{page}{0}
\setcounter{tocdepth}{2}

\newpage

\renewcommand{\thefootnote}{\arabic{footnote}}
\setcounter{footnote}{0}

\linespread{1.1}
\parskip 4pt



\section{Introduction}

AdS/CFT methods have been successfully used in order to calculate transport in condensed matter models, though the 
particular functional behaviours usually are either different than, or more general than ones obtained in real materials, and 
one must phenomenologically (ad-hoc) fix parameters and/or functions to obtain a fit. This so-called ``AdS/CMT'' method 
is therefore viewed best as a phenomenological one, and must therefore be considered within the most general holographic 
model available. One is led to consider a system of gravity plus Abelian vector field, plus a scalar that defines the kinetic 
functions appearing in the Lagrangian. 

Transport in such systems has been considered in many papers, but here we will be mostly interested in the methods 
used in \cite{Blake:2015ina,Donos:2015bxe,Donos:2014cya,Banks:2015wha,Donos:2014uba,Donos:2017mhp,Erdmenger:2016wyp}. The question we want to ask is, what is the effect of S-duality on this bulk holographic theory on  the transport
coefficients for the holographic dual field theory? The S-duality should correspond to particle-vortex duality in 
the boundary \cite{Murugan:2014sfa,Alejo:2019hnb}. We will not consider the effect of quantum 
gravitational corrections to the bulk gravity action (those have been addressed in \cite{Alejo:2019hnb}). Since we 
are after the effect of S-duality, we will consider a vector action that involves both $F_{\mu\nu}$ and its dual $\tilde 
F_{\mu\nu}$. Transport will be calculated using three different methods, a standard membrane paradigm type 
method at the horizon for nonextremal black holes, the entropy function formalism for extremal black holes (considered in 
conjunction with a $T\rightarrow 0$ limit of the previous formalism), 
and the formalism of (fluid) Stokes equations in the case of one-dimensional lattices. The last formalism is also 
considered in the $T\rightarrow 0$ limit and then generalized, in order to take advantage of a supergravity-inspired 
model for which we can apply the same entropy function formalism.  In all of these 3 formalisms, we consider the 
effect of S-duality of the model on the transport coefficients.   

The paper is organized as follows. In section 2 we define the model, the behaviour at the black hole horizon, and we add 
magnetization currents in the presence of external magnetic fields, studying the resulting thermodynamics. 
In section 3 we calculate electric and thermal transport in this model, calculating the resulting transport coefficients, and 
study the effet of S-duality on them.
In section 4 we use the entropy function formalism, for extremal black holes, to calculate the transport coefficients, in the 
corresponding limit of the formulas from section 3, as a function of only the charges of the dual black hole. We 
also explore a subtlety of S-duality in this limit. 
In section 5 we consider the formalism of Stokes equations to calculate the transport coefficients, and apply it to 
one-dimensional lattices. S-duality in this case is also explored. 
In section 6, we apply the results of section  5 to a supergravity-inspired model, by generalizing the formulas for 
transport coefficients and using the entropy function formalism. In section 7 we conclude.

\section{AdS/CMT model and black hole horizon data}

\subsection{Model and black hole horizon}

Following the logic from \cite{Blake:2015ina}, we consider 3+1 dimensional
gravity coupled to an Abelian vector field $A_\mu$, 
with both a Maxwell and a ``theta'' (topological) term, and kinetic functions $Z(\phi),W(\phi)$ 
defined by a scalar ``dilaton'' $\phi$, which has some potential $V(\phi)$. For more generality, in order to break translational 
invariance in one or two spatial directions, we can consider also two more scalar ``axions'' $\chi_1,\chi_2$
that have VEV linear in the coordinates $x,y$ and kinetic function $\Phi(\phi)$. The action is therefore 
\bea
S=\int d^{4}x \sqrt{-g}\left[\frac{1}{16\pi G_{N}}\left(R-\frac{1}{2}[(\partial \phi)^{2}+\Phi(\phi)\left((\partial\chi_{1})^{2}+(\partial\chi_{2})^{2}\right)]-V(\phi)\right)\right. \nonumber \\
\left. -\frac{Z(\phi)}{4g_{4}^{2}}F^{2}_{\mu\nu}-W(\phi)F_{\mu\nu}\tilde{F}^{\mu\nu}\right], \label{action}
\eea
where we note the addition of the topological term with coefficient function $W(\phi)$ as compared to  \cite{Blake:2015ina},
in order to be able to study S-duality consistently.

Here the field strength $F_{\mu\nu}$ and the dual field strength $\tilde F_{\mu\nu}$ are defined as
\be 
F_{\mu\nu}=\partial_{\mu}A_{\nu}-\partial_{\nu}A_{\mu}, \,\,\, \tilde{F}^{\mu\nu}=\frac{1}{2}\frac{\tilde{\epsilon}^{\mu\nu\rho\sigma}}{\sqrt{-g}}F_{\rho\sigma}\;,
\ee
while the linear axion background solution is 
\be 
\chi_{1}=k_{1}x, \,\,\, \chi_{2}=k_{2}y. 
\ee

We are interested in models with a holographic dual, so the solutions we want to use must be asymptotically AdS, meaning 
that the scalar potential must have an AdS solution, so 
\begin{equation}
V(0) = -\frac{6}{L^2},~V'(0)=0.\label{PotentialCond} 
\end{equation}

The equations of motion for the gravity and the gauge field are 
\be 
R_{\mu\nu}=\frac{1}{2}\partial_{\mu} \phi\partial_{\nu} \phi+\frac{1}{2}g_{\mu\nu}V(\phi)+\frac{(16\pi G_{N})}{4g_{4}^{2}}Z(\phi)\left(2F_{\mu\lambda}{F_{\nu}}^{\lambda}-\frac{1}{2}g_{\mu\nu}F_{\rho\sigma}F^{\rho\sigma}\right)\label{einstein}, \ee
\be \frac{1}{\sqrt{-g}}\partial_{\mu}\left[\sqrt{-g}\left(\frac{Z(\phi)}{g_{4}^{2}}F^{\mu\nu}+4W(\phi)\tilde{F}^{\mu\nu}\right)\right]=0. \label{gauge}
\ee

We have not written the equation of motion for the scalar dilaton $\phi$ (not for the linear dilatons $\chi_1,\chi_2$), 
but we assume that it has solutions that 
asymptotically satisfy the condition (\ref{PotentialCond}). 

For the {\em isotropic case} (with $\chi_1=\chi_2=0$), 
the background metric plus gauge field solutions we consider are of the type
\bea
ds^{2}&=&-U\, dt^{2}+U^{-1}dr^{2}+e^{2V}(dx^{2}+dy^{2})\cr
A&=&a(r) dt-Bydx, \label{metdyonic}
\eea
where $U=U(r), V=V(r)$ (note that $V(r)$ is a factor in the metric and $V(\phi)$ is the scalar potential). 

The solutions of interest must have a  temperature $T$, since the dual field theory, whose transport we want to calculate, 
must have the same. That means that we are interested in black hole solutions that asymptote to AdS space, and have 
event horizons at $r=r_H$. Near it, the background fields are expanded as
\bea
U(r) & \simeq& U(r_H)+(r-r_H)U'(r_H)+{\cal O}((r-r_H)^2)=4\pi T(r-r_{+})+... , \cr
a(r) & \simeq& a_{H}(r-r_{H})+... , \cr
V(r) &\simeq& V(r_{H})+... , \cr
\phi & \simeq& \phi_{H}+... , \label{Fieldexp}
\eea
where we assume $U(r_H)=0$ for the existence of the event horizon and $U'(r_H)\neq 0$ for a non-extremal solution.

The near-horizon metric for the {\em non-extremal} black hole then becomes (in the extremal case $U'(r_H)=0$ also, 
and we need to go to the next order)
\be 
ds^{2}\simeq 
-(r-r_{H})U'(r_{H})dt^{2}+\frac{1}{(r-r_{H})U'(r_{H})}dr^{2}+e^{2V(r_{H})}(dx^{2}+dy^{2})\;,\label{apmet}
\ee
which is of the type of two-dimensional Rindler spacetime times $\mathbb{R}^2$. The surface gravity is $\kappa=\pm 
U'(r_H)/2$, the corresponding temperature (in units where $\hbar=k_B=1$) being
\be
T=\frac{\kappa}{2\pi}=\frac{U'(r_H)}{4\pi}.
\ee

With the change of coordinates $r-r_H=U'(r_H)z^2/4$, the Rindler space part of the metric is 
\be
ds^{2}=-(\kappa z)^{2}dt^{2}+dz^{2}. 
\ee

The near-horizon solution admits 3 scaling symmetries,
\be 
t\rightarrow \lambda t, \,\, \kappa\rightarrow \lambda^{-1} \kappa, \ee
\be t\rightarrow \chi^{-1} t,\,\,\, (r-r_{H})\rightarrow \chi (r-r_{H}), \,\, U'(r_{H})\rightarrow \chi U'(r_{H}),  \ee
\be e^{V(r_{H})}\rightarrow \xi e^{2V(r_{H})}, \,\, x\rightarrow \xi^{-1} x, \,\,\, y \rightarrow \xi^{-1} y.  
\ee

\subsection{Magnetizations and thermodynamics}

In the next section we will study electric and thermal (heat) transport, but it is interesting to consider it in the presence 
of a magnetic field, for generality of the treatment. In this case however, it is known that there is an extra 
magnetic contribution to the electric and heat currents $\vec{J}$ and $\vec{Q}$, depending on the magnetization density
$M$  and energy magnetization density $M_E$, and being of the Hall (off-diagonal) type, 
\bea
J_i^{\rm (mag)}&=& \frac{M}{T}\epsilon_{ij}\nabla_j T\cr
Q_i^{\rm (mag)}&=&M \epsilon_{ij}E_j+\frac{2(M_E-\mu M)}{T}\epsilon_{ij}\nabla_j T.
\eea
Here both $M$ and $M_E$ are defined for the boundary 2+1 dimensional field theory as responses of the theory to a 
source that changes the fields, and $M_Q=M_E-\mu M$ is called heat magnetization density. 
For a source $A_x^{(0)}=-By$, giving a magnetic field $B$ in 2+1 dimensions, the 
magnetization density is (minus) the variation of the (density of the) Euclidean action with respect to $B$, 
\be
M=-\frac{1}{Vol}\frac{\d S_E}{\d B}\;,
\ee
whereas the energy magnetization density is the same thing if we apply a change in the (Minkowski) metric of the field 
theory, with source $\delta g^{(0)}_{tx}=-B_1y$, and differentiate with respect to $B_1$, 
\be
M_E=-\frac{1}{Vol}\left.\frac{\d S_E}{\d B_1}\right|_{B_1=0}.\label{enmag}
\ee
Here the Euclidean action in the bulk is 
\be  
S_{E}=\int d^{4}x \sqrt{g}\left[\frac{1}{16\pi G_{N}}\left(R+\frac{1}{2}(\partial \phi)^{2}
+V(\phi)\right)+\frac{Z(\phi)}{4g_{4}^{2}}F^{2}-W(\phi)F_{\mu\nu}\tilde{F}^{\mu\nu}\right]. \label{euc}
\ee

The effect of this source on the boundary is to introduce a $\delta g^{(0)}_{tx}=-U(r)B_1y$ in the bulk, and by 
consistency of the equations of motion, we need also to add to $A$ a term $(a(r)-\mu)B_1ydx$, where $\mu$ is the 
boundary chemical potential, obtaining a modified background solution of ($\chi_1=k_1x, \chi_2=k_2y, \phi=\phi(r)$ and)
\bea
A_{t}&=&a(r), \,\,\, A_{x}=-By+(a(r)-\mu)B_{1}y, \cr
ds^{2}&=&-U(r)(dt+B_{1}ydx)^{2}+\frac{dr^{2}}{U(r)}+e^{2V(r)}(dx^{2}+dy^{2}). 
\eea

The inverse metric is then (in $t,r,x,y$ space)
\be 
g^{\mu\nu}=\left[
\begin{array}{cccc}
 B_{1}^2 e^{-2 V} y^2-\frac{1}{U} & 0 & -B_{1} e^{-2 V} y & 0 \\
 0 & U & 0 & 0 \\
 -B_{1} e^{-2 V} y & 0 & e^{-2 V} & 0 \\
 0 & 0 & 0 & e^{-2 V}
\end{array}
\right]
\ee

After some algebra, we obtain the Maxwell field Euclidean action in the bulk, on this ansatz, as 
\bea 
S_{E}^{\text{Maxwell}}=\int d^{4}x\left[\frac{Z(\phi)e^{2V}}{4g_{4}^{2}}\left(2(a')^{2}-
2e^{-4V}[-B+(a(r)-\mu)B_{1}]^{2}\right)\right. \nonumber \\
\left. -4W(\phi)a'(r)(-B+(a(r)-\mu)B_{1})\right].
\eea

We then obtain the magnetization density, energy magnetization density, and heat magnetization density as
\bea
M&=&-\frac{1}{V}\frac{\partial S_{E}}{\partial B}=\int_{r_{H}}^{\infty}dr 
\left(\frac{e^{-2V}Z(\phi)B}{g_{4}^{2}}-4W(\phi)a'(r) \right)\label{magdensity}\\
M_{E}&=&\int_{r_{H}}^{\infty}dr \left(\frac{e^{-2V}Z(\phi)B}{g_{4}^{2}}-4W(\phi)a'(r)\right)(\mu -a(r)) \\
M_{Q}&=&M_{E}-\mu M=- \int_{r_{H}}^{\infty}dr \left(\frac{e^{-2V}Z(\phi)B}{g_{4}^{2}}-4W(\phi)a'(r)\right)a(r). \label{heatmag}
\eea

\section{Transport and S-duality}

In this section we calculate electric and heat transport for the background solutions from the previous section, in order to  study 
the effect of S-duality on it. 

We add perturbations and electrical and thermal gradient 
sources to the background solution of the previous section, with the same notation as in 
\cite{Blake:2015ina,Donos:2014uba}, in the presence of a magnetic field $B$, but at $B_1=0$. The electric field
perturbation is sourced by a boundary electric field $E$ and thermal gradient $\frac{1}{T}\nabla_i T$ of 
\be
E_i=E\delta_{ix}\;,\;\;\; 
\frac{1}{T}\nabla_i T=\xi \delta_{ix}.
\ee
This results in a extra gauge field term in the bulk of $(-E+\xi a(r))tdx$ and an extra metric term of $\delta g^{(0)}_{tx}
=-\xi t U$, so adding relevant perturbations we 
obtain the perturbed ansatz (the diagonal metric and $A_t$ are unperturbed)
\bea
A_t&=&a(r)\cr
A_{x}&=&-By+(-E+\xi a(r))t+\delta A_{x}(r) \nonumber \\
A_{y}&=&\delta A_{y}(r)\nonumber \\
g_{tx}&=&-\xi t U +e^{2V}\delta h_{tx}(r) \nonumber \\
g_{ty}&=&e^{2V}\delta h_{ty}(r) \nonumber \\
g_{rx}&=&e^{2V}\delta h_{rx}(r) \nonumber \\
g_{ry}&=&e^{2V}\delta h_{ry}(r) \nonumber \\
\chi_{1}&=&kx+\delta \chi_{1}(r) \nonumber \\
\chi_{2}&=&ky+\delta \chi_{2}(r).
\label{allpert}
\eea
Note that the logic is that the sources $E,B,\xi$ are small, and they in turn generate the perturbations $\delta h_{\mu\nu}$, 
solved to linear order from the Einstein's equations, as a function of the sources (linear response theory).

Putting an explicit $\epsilon$ in the perturbation matrix (for Mathematica computation reasons), the metric and its 
inverse to order $\epsilon$, in matrix form (for a space $t,r,x,y$), and the field strength components, are
\bea
g&=&\left(
\begin{array}{cccc}
 -U & 0 & e^{2 V} \delta h_{tx} \epsilon -t U
   \epsilon  \xi  & e^{2 V} \delta h_{ty} \epsilon
    \\
 0 & \frac{1}{U} & e^{2 V} \delta h_{rx} \epsilon 
   & e^{2 V} \delta h_{ry} \epsilon  \\
 e^{2 V} \delta h_{tx} \epsilon -t U \epsilon  \xi
    & e^{2 V} \delta h_{rx} \epsilon  & e^{2 V} &
   0 \\
 e^{2 V} \delta h_{ty} \epsilon  & e^{2 V}
   \delta h_{ry} \epsilon  & 0 & e^{2 V}
\end{array}
\right),\cr
g^{-1}&=&
\left(
\begin{array}{cccc}
 -\frac{1}{U} & 0 & \epsilon  \left(\frac{\delta
   h_{tx}}{U}-e^{-2 V} t \xi \right) &
   \frac{\delta h_{ty} \epsilon }{U} \\
 0 & U & -U \delta h_{rx} \epsilon  & -U
   \delta h_{ry} \epsilon  \\
 \epsilon  \left(\frac{\delta h_{tx}}{U}-e^{-2 V}
   t \xi \right) & -U \delta h_{rx} \epsilon  &
   e^{-2 V} & 0 \\
 \frac{\delta h_{ty} \epsilon }{U} & -U
   \delta h_{ry} \epsilon  & 0 & e^{-2 V}
\end{array}
\right)\label{metwithpert}\\
F_{rt}&=&a^{\prime}, \nonumber \\ 
F_{tx}&=&\epsilon(-E+\xi a), \nonumber \\
F_{xy}&=&B, \nonumber\\
F_{rx}&=&\epsilon\xi a^{\prime}t+\epsilon\delta A^{\prime}_x, \nonumber \\
F_{ry}&=&\epsilon\delta A^{\prime}_y.
\eea

The gauge field equations, $x$ and $y$ components, are 
\bea
0&=&\partial_t\left(\frac{\sqrt{-g}Z(\phi)}{g^2_4}F^{tx}+4\sqrt{-g}W(\phi)\tilde{F}^{tx}\right)
+\partial_r \left(\frac{\sqrt{-g}Z(\phi)}{g^2_4}F^{rx}+4\sqrt{-g}W(\phi)\tilde{F}^{rx}\right)\nonumber\\
&+& \partial_y\left(\frac{\sqrt{-g}Z(\phi)}{g^2_4}F^{yx}+4\sqrt{-g}W(\phi)\tilde{F}^{yx}\right)\label{bndry1}\\
0&=&\partial_t\left(\frac{\sqrt{-g}Z(\phi)}{g^2_4}F^{ty}+4\sqrt{-g}W(\phi)\tilde{F}^{ty}\right)+\partial_r 
\left(\frac{\sqrt{-g}Z(\phi)}{g^2_4}F^{ry}+4\sqrt{-g}W(\phi)\tilde{F}^{ry}\right)\nonumber\\
&+& \partial_x\left(\frac{\sqrt{-g}Z(\phi)}{g^2_4}F^{xy}+4\sqrt{-g}W(\phi)\tilde{F}^{xy}\right)\;,\label{bndry2}
\eea
and become on the ansatz to leading order
\bea
0&=&-\partial_t\left[\frac{1}{g_{4}^{2}}\left(a^{\prime}Ze^{2V}\delta h_{rx}+\frac{\xi aZ}{U}
+\frac{ZB}{U}\delta h_{ty}-\frac{EZ}{U} \right)+4W\delta A_{y}^{\prime}\right]\nonumber\\
&=&-\partial_r \left(\frac{\sqrt{-g}Z(\phi)}{g^2_4}F^{rx}+4\sqrt{-g}W(\phi)\tilde{F}^{rx}\right)\;, \label{xeqf}\\
&&\partial_t \left[ -\frac{1}{g_{4}^{2}}\left(a^{\prime}e^{2V}Z\delta h_{ry}-\frac{BZ}{U}\delta h_{tx}
+ZBe^{-2V}t\xi \right)+4W(\xi a^{\prime}t+\delta A^{\prime}_x)\right]\nonumber \\
&=&-\frac{Z}{g_{4}^{2}}\xi e^{-2V}B+4\xi Wa^{\prime}\nonumber \\
&=&\partial_r\left(\frac{Z}{g_{4}^{2}}\sqrt{-g}F^{yr}+4\sqrt{-g}W\tilde{F}^{yr}\right)\label{yeqf}.
\eea

\subsection{Electric current, conductivity and thermoelectric coefficients}

The calculation of the transport coefficients of the dual field theory at the horizon of the black hole relies on the 
membrane paradigm idea, first present in the calculation of \cite{Iqbal:2008}, that the quantities appearing in the 
currents are independent of the radial position $r$, so instead of calculating them at the boundary at $r\rightarrow \infty$, 
like the AdS/CFT prescription dictates, we can calculate them at the horizon. But if it is the case that the currents do depend 
on $r$, like in \cite{Blake:2015ina}, we must redefine them, and find quantities that {\em can} be calculated at the 
horizon, being $r$ independent. 

The standard (and total) current, defined according to \cite{Iqbal:2008} (see also \cite{Lopez-Arcos:2013uga}), would be 
\be
j^{i{\rm (tot)}}=\frac{\delta S}{\delta \d_r A_i}=\frac{Z(\phi)}{g_4^2}\sqrt{-g}F^{ir}+4\sqrt{-g} W(\phi)\tilde F^{ir}\;,
\ee
where $S$ is the full bulk action. But we note that, because of (\ref{yeqf}), the $y$ component of the gauge field 
equation is not $r$-independent, so cannot be calculated at the horizon. 

We must calculate instead the modified currents (or fluxes) defined as 
\bea
{\cal J}^x&=&\frac{Z(\phi)}{g_{4}^{2}}\sqrt{-g}F^{xr}+4\sqrt{-g}W(\phi)\tilde{F}^{xr},\nonumber\\
{\cal J}^{y}&=&\frac{Z(\phi)}{g_{4}^{2}}\sqrt{-g}F^{yr}+4\sqrt{-g}W(\phi)\tilde{F}^{yr}-\xi M(r)\;,\label{Jx}
\eea
which are now independent of $r$, since $M(r)$ is a position-dependent magnetization density
given by \eqref{magdensity}, only integrated up to $r$ only instead 
of all the way to $\infty$, so that $\d_r$ on it gives the bracket in (\ref{magdensity}) as the extra term in (\ref{yeqf}).

Explicitly, we obtain the fluxes
\bea
{\cal J}^x&=&-\epsilon \frac{Z}{g_{4}^{2}}a^{\prime}e^{2V}\delta h_{tx}
-\epsilon \frac{Z}{g_{4}^{2}}U\delta A^{\prime}_x-\epsilon \frac{Z}{g_{4}^{2}}UB\delta h_{ry}\cr
{\cal J}^y&=&-\frac{Z}{g_{4}^{2}}U\delta A^{\prime}_y-\frac{Z}{g_{4}^{2}}e^{2V}a^{\prime}\delta h_{ty}
+\frac{Z}{g_{4}^{2}}BU\delta h_{rx}+4W(-E+\xi a)-\xi M(r)\;,
\eea
which can then be evaluated at any $r$, including $r_H$ (the horizon). 

The important observation is that, while $\d_r {\cal J}^i=0$, so we can calculate them at the horizon, at infinity
$M(r)=M(\infty)=M$ is just the magnetization, so we just subtract the magnetization currents from the total currents, 
obtaining the usual transport currents, from which we can calculate the conductivity and thermoelectric coefficients,
\be
{\cal J}^i(r=r_H)={\cal J}^i(r\rightarrow \infty)=j^{i{\rm (tot)}}-\xi M=j^i.
\ee

The advantage of being able to calculate at the horizon is that we can impose the conditions of regularity at the
horizon (remember that $E_i=E\delta_{ix}$ and $\xi_i=\xi \delta_{ix}$)
\bea
\delta A_i&=&-\frac{E_i}{4\pi T}\ln(r-r_{H})+{\cal O}(r-r_{H}),\nonumber\\
\delta \chi_i&=&{\cal O}((r-r_{H})^0),\nonumber\\
\delta h_{ti}&=&U\delta h_{ri}-\frac{\xi_iU}{4\pi e^{2V}T}\ln(r-r_{H})+{\cal O}(r-r_{H}),\label{Regul}
\eea
and moreover, since $M(r)$ is an integral from $r_H$ to $r$, it vanishes at the horizon, simplifying the result.
Using (\ref{Regul}), we obtain that the fluxes at the horizon, equaling the transport currents, are
\bea
j^x={\cal J}^x(r_H)&=&\left. \frac{Z}{g_{4}^{2}}E_x-\frac{Z}{g_{4}^{2}}e^{2V}a^{\prime}\delta h_{tx}
-\frac{Z}{g_{4}^{2}}B\delta h_{ty}\right|_{r_{H}}, \nonumber\\
j^y={\cal J}^y(r_H)&=&\left.\frac{Z}{g_{4}^{2}}E_y-e^{2V}Za^{\prime}\delta h_{ty}
+\frac{Z}{g_{4}^{2}}B\delta h_{tx}-4W(E+\xi a)\right|_{r_{H}}. \label{curr}
\eea

As we see, it remains to solve for $\delta h_{ti}$ using the Einstein's equations, as a function of the external 
sources $E,B,\xi$ (linear response theory). Since the topological term with $W(\phi)$ doesn't contribute to Einstein's equations
(it is independent of the metric), the linearized Einstein's equations are the same as in \cite{Blake:2015ina}, namely
\bea  
U(e^{4V}\delta h'_{tx})'-\left(\frac{2\kappa^{2}}{g_{4}^{2}}ZB^{2}+e^{2V}k^{2}\Phi\right)\delta h_{tx}+
\frac{2\kappa^{2}}{g_{4}^{2}}ZBUe^{2V}a'\delta h_{ty}&=&-\frac{2\kappa^{2}}{g_{4}^{2}}Ze^{2V}a'\delta
 a_{x}', \nonumber \\
U(e^{4V}\delta h'_{ty})'-\left(\frac{2\kappa^{2}}{g_{4}^{2}}ZB^{2}+e^{2V}k^{2}\Phi\right)\delta h_{ty}-
\frac{2\kappa^{2}}{g_{4}^{2}}ZBUe^{2V}a'\delta h_{tx}&=&-\frac{2\kappa^{2}}{g_{4}^{2}}Ze^{2V}a'
\delta a_{y}'\cr
&& +\frac{2\kappa^{2}}{g_{4}^{2}}ZB(-E+\xi a).\cr
&&
\eea
Using the regularity conditions at the horizon (\ref{Regul}), we obtain 
\bea
\label{Eins}
\left(\frac{2\kappa^{2}}{g_{4}^{2}}ZB^2+e^{2V}k^2\Phi\right)\delta h_{tx}
-\frac{2\kappa^{2}}{g_{4}^{2}}ZBe^{2V}a^{\prime}\delta h_{ty}&=&-\frac{2\kappa^{2}}{g_{4}^{2}}Z
e^{2V}a^{\prime}E+e^{2V}U^{\prime}\xi,\nonumber \\
\left(\frac{2\kappa^{2}}{g_{4}^{2}}ZB^2+e^{2V}k^2\Phi\right)\delta h_{ty}
-\frac{2\kappa^{2}}{g_{4}^{2}}ZBe^{2V}a^{\prime}\delta h_{tx}&=&\frac{2\kappa^{2}}{g_{4}^{2}}ZBE\;,
\eea
and after some algebra we solve the $\delta h_{ti}$ graviton fluctuations in terms of the sources $E,B,\xi$ as 
\bea
\delta h_{tx}&=&\frac{\frac{2\kappa^{2}}{g_{4}^{2}}Ze^{4V}a^{\prime}k^2\Phi}{(\frac{2\kappa^{2}}
{g_{4}^{2}}ZB^2+e^{2V}k^2\Phi)^2+(\frac{2\kappa^{2}}{g_{4}^{2}}Z)^{2}B^2e^{4V}a^{\prime 2}}E\cr
&&+\frac{(\frac{2\kappa^{2}}{g_{4}^{2}}ZB^2+e^{2V}a^{\prime}k^2\Phi)e^{2V}U^{\prime}\xi}
{(\frac{2\kappa^{2}}{g_{4}^{2}}ZB^2+e^{2V}k^2\Phi)^2+(\frac{2\kappa^{2}}{g_{4}^{2}}Z)^{2}
B^2e^{4V}a^{\prime 2}},\nonumber\\
\delta h_{ty}&=&\frac{\frac{2\kappa^{2}}{g_{4}^{2}}ZB}{(\frac{2\kappa^{2}}{g_{4}^{2}}ZB^2+e^{2V}k^2\Phi)}E
-\frac{2\kappa^{2}}{g_{4}^{2}}ZBe^{2V}a^{\prime}\frac{e^{2V}U^{\prime}\xi}{[(\frac{2\kappa^{2}}
{g_{4}^{2}}ZB^2+e^{2V}k^2\Phi)^2+(\frac{2\kappa^{2}}{g_{4}^{2}}Z)^{2}B^2e^{4V}a^{\prime 2}]}\nonumber \\
&\phantom{.}&-\frac{\frac{2\kappa^{2}}{g_{4}^{2}}ZBe^{2V}a^{\prime}}{(\frac{2\kappa^{2}}{g_{4}^{2}}
ZB^2+e^{2V}k^2\Phi)[(\frac{2\kappa^{2}}{g_{4}^{2}}ZB^2+e^{2V}k^2\Phi)^2+(\frac{2\kappa^{2}}
{g_{4}^{2}}Z)^{2}B^2e^{4V}a^{\prime 2}]}\times\cr
&&\times e^{4V}\frac{2\kappa^{2}}{g_{4}^{2}}Za^{\prime}k^2
\Phi E.\label{grav}
\eea
We can now replace the fluctuations (\ref{grav}) in the currents (\ref{curr}), use the fact that $U'(r_H)=4\pi T$
(meaning that $U'(r_H) \xi=4\pi T\xi$) 
and separate the terms according to the 
sources $E$ and $\xi$, via
\be
j^i=\sigma_{ij}E^j-\alpha_{ij}(\nabla T)_j=\sigma_{ix}E-\a_{ix}T\xi.
\ee
From this, we can identify directly the conductivities and thermoelectric coefficients as
\bea
\sigma_{xx}&=&\left.\frac{e^{2V}k^{2}\Phi(2\kappa_{4}^{2}g_{4}^{4}\rho^{2}+2\kappa_{4}^{2}B^{2}Z^{2}
+g_{4}^{2}Ze^{2V}k^{2}\Phi)}{4\kappa_{4}^{4}g_{4}^{4}B^{2}\rho^{2}+(2\kappa_{4}^{2}B^{2}Z
+g_{4}^{2}e^{2V}k^{2}\Phi)^{2}}\right|_{r_H}, \cr
\sigma_{xy}&=&\left.4\kappa_{4}^{2}B\rho\frac{\kappa_{4}^{2}g_{4}^{4}\rho^{2}+\kappa_{4}^{2}B^{2}Z^{2}
+g_{4}^{2}Ze^{2V}k^{2}\Phi}{4\kappa_{4}^{4}g_{4}^{4}B^{2}\rho^{2}+(2\kappa_{4}^{2}B^{2}Z
+g_{4}^{2}e^{2V}k^{2}\Phi)^{2}}-4W\right|_{r_H},  \label{elconddxy}\\
\alpha_{xx}&=&\left.\frac{2\kappa_{4}^{2}g_{4}^{4}s\rho e^{2V}k^{2}\Phi}{4\kappa_{4}^{4}g_{4}^{4}
B^{2}\rho^{2}+(2\kappa_{4}^{2}B^{2}Z+g_{4}^{2}
e^{2V}k^{2}\Phi)^{2}}\right|_{r_H}, \cr
\alpha_{xy}&=&\left. 2\kappa_{4}^{2}sB\frac{2\kappa_{4}^{2}g_{4}^{4}\rho^{2}+2\kappa_{4}^{2}B^{2}Z^{2}
+g_{4}^{2}Ze^{2V}k^{2}\Phi}{4\kappa_{4}^{4}g_{4}^{4}B^{2}\rho^{2}+(2\kappa_{4}^{2}B^{2}Z
+g_{4}^{2}e^{2V}k^{2}\Phi)^{2}}\right|_{r_H}\;,\label{etconddxy} 
\eea
which have been expressed in terms of the boundary charge density $\rho=-Z e^{2V}a'$ and the entropy density 
(Hawking formula) $s=4\pi e^{2V(r_H)}$.

We note that the only new contribution is from the $W$ term in the Hall conductivity $\sigma_{xy}$.

\subsection{Heat current and heat conductivity}

We move on to the heat current, considered (also with a topological term) from the point of view of 
transport coefficients in the presence of magnetization in \cite{Donos:2017mhp}. The heat current itself was defined 
in \cite{Donos:2014cya}.

The (total) heat current is obtained from the energy-momentum tensor by subtracting the electric current, 
\be
Q^{{\rm (tot)}i}=T^{{\rm (tot)} i0}-\mu J^{{\rm (tot)}i}.
\ee
But in \cite{Donos:2014cya,Donos:2017mhp} it was noticed that the result one obtains from this equals, at $r\rightarrow 
\infty$, the flux
\be
Q^{{\rm (tot)}i}=\sqrt{-g} G^{ri}\;,
\ee
where the bulk 2-form $G^{\mu\nu}$ is defined through 
\be
G^{\mu\nu}=-2\nabla^{[\mu}k^{\nu]}-Zk^{[\mu}F^{\nu]\rho}A_{\rho}
-\frac{1}{2}(\psi-2\theta)H^{\mu\nu},
\label{btf1}
\ee
where
\be
H^{\mu\nu}\equiv Z(\phi)F^{\mu\nu}+4g_4^2W(\phi)\tilde F^{\mu\nu}\;,
\ee
and where $k^\mu$ is the vector $\d_t$. However, more generally, an arbitrary vector satisfying 
$\nabla_\mu k^\mu=0$ will also satisfy (as can be easily checked) the general property
\be
\nabla_{\mu}(\nabla^{[\mu}k^{\nu]})=\nabla_{\mu}(\nabla^{(\mu}k^{\nu)})-{R^{\nu}}_{\mu}k^{\mu},\label{qKill}
\ee
which is the only one we need. The functions $\psi$ and $\theta$ are defined by the relations
\bea
\nabla_{\rho}\psi &=&({\mathcal L}_{k}A)_{\rho}=k^{\mu}\partial_{\mu}A_{\rho}
+A_{\mu}\partial_{\rho}k_{\mu}, \label{covpsi} \\
\nabla_{\rho}\theta & =& k^{\mu}F_{\mu\rho}
-\frac{1}{2}\xi_{\rho}k^{\mu}A_{\mu}.\label{covtheta}
\eea

After some involved algebra, we obtain 
\bea
\nabla_{\mu}G^{\mu\nu}&=&Vk^{\nu}-2\nabla_{\mu}(\nabla^{(\mu}k^{\nu)})+\frac{1}{2}ZF^{\nu\mu}s_{\mu}
-\frac{Z}{2}A_{\rho}(\mathcal{L}_{k}F)^{\nu\rho}\nonumber \\
&&-2g_{4}^{2}(\partial_{\mu} W)\tilde{F}^{\mu\rho}A_{\rho}k^{\nu}
- 2g_{4}^{2}W \tilde{F}^{\mu\nu}\nabla_{\mu}(\psi-2\theta)\;, \label{hope}
\eea
where 
\bea
(\mathcal{L}_kF)^{\nu\rho}&=&k^{\mu}\nabla_{\mu}F^{\nu\rho}-\nabla_{\mu}k^{\nu}F^{\mu\rho}
-\nabla_{\mu}k^{\rho}F^{\nu\mu}\cr
s_{\mu}&\equiv &k^{\nu}F_{\nu\mu}-\nabla_{\mu}\theta\;,
\eea
and where we define {\em on-shell} $V$ by 
\be
2{R^\mu}_\nu k^\nu=V k^\mu.
\ee
We also calculate, in the $A_r=0$ gauge and in the background (no fluctuations), and using the fact that $k^\mu=(\d_t)^\mu$,
\bea
\int dx^{\rho}\nabla_{\rho}(\psi-2\theta)&=&\int dx^{\rho}k^{\mu}\partial_{\mu}A_{\rho}
+\int dx^{\rho}A_{\mu}\nabla_{\rho}k^{\mu}-2\int dx^{\rho}k^{\mu}F_{\mu\rho}+\int dx^{\rho}\xi_{\rho}i_k A\cr
&=&Ex+2a.
\eea

Further,
\bea
G^{ri}&=&-\nabla^rk^i+\nabla^ik^r-Z(\Phi)k^{[r}F^{i]\sigma}A_{\sigma}-\frac{1}{2}(2a(r)+Ex)H^{ri}\cr
\phantom{..}&=& -g^{r\alpha}\Gamma^{i}_{\alpha t}+g^{i\alpha}\Gamma^r_{\alpha t}
-\frac{1}{2}(2a(r)+Ex)H^{ri}\;,
\eea
so,  after some calculations {\em in the presence of fluctuations}, we find that {\em at $r\rightarrow \infty$}, when 
$a(r)$ dominates over $Ex$, we have
\be
-Q^{{\rm (tot)}i}=-\sqrt{-g} G^{ri}=U^2\left(\frac{e^{2V}\delta h_{ti}}{U}\right)^{\prime}
+a(r)\sqrt{-g}(Z(\phi)F^{ri}+4g_{4}^{2}W(\phi)\tilde{F}^{ri}).\label{heattot}
\ee
Note that
\bea
F^{rx}&=&\epsilon(a'\delta h_{tx}+Ue^{-2V}\delta A_x' +U e^{-2V}B \delta h_{ry})\cr
F^{ry}&=&\epsilon(a'\delta h_{ty}+Ue^{-2V}\delta A_y'-Ue^{-2V}B \delta h_{rx})\cr
\sqrt{-g}\tilde F^{rx}&=&0\cr
\sqrt{-g}\tilde F^{ry}&=&-\epsilon(-E+\xi a).
\eea

However, from (\ref{hope}), we find that 
\be
\sqrt{-g}\nabla_\mu G^{\mu i}=\d_\mu (\sqrt{-g}G^{\mu i})\neq 0\;,
\ee
and it equals zero only in the absence of thermal fluctuations (which we are interested in).
If it would be true, we would have that the linearized fluxes $\sqrt{-g}G^{ri}$ would be independent of $r$, and 
could be evaluated at the horizon. 

As it is, we obtain from evaluating (\ref{hope}) the {\em modified} conservation laws, 
\bea
\partial_r(\sqrt{-g}G^{rx})&=&-\partial_t(\sqrt{-g}G^{tx})-\partial_y(\sqrt{-g}G^{yx}),\nonumber\\
\partial_r(\sqrt{-g}G^{ry})&=&-\partial_t(\sqrt{-g}G^{ty})-\partial_x(\sqrt{-g}G^{xy})+
\sqrt{-g}H^{xy}a(r).
\eea

Moreover, we calculate
\bea
G^{tx}&=&-g^{tt}\Gamma^x_{tt}+g^{xr}\Gamma^t_{rt}-\frac{1}{2}ZF^{xt}A_t-\frac{1}{2}ZF^{xy}A_y
-\frac{1}{2}(2a+Ex)(ZF^{tx}+4g_{4}^{2}WF_{yr})\cr
G^{xy}&=&-\frac{1}{2}(2a+Ex)(Ze^{-4V}B-4g_{4}^{2}e^{-2V}a')=-\frac{1}{2}(2a+Ex)H^{xy}\cr
G^{ty}&=&-\frac{U'}{U}\delta h_{ry}-\frac{1}{2}Z(e^{-4V}B\xi t+\cdot\cdot\cdot)a\cr
&&-\frac{1}{2}Z\left[\left(-e^{-4V}B+\frac{\delta h_{ty} e^{-2V}}{U}(-E+\xi a)\right)(-E+\xi a)t+...\right]\nonumber\\
&\phantom{.}&-\frac{1}{2}(2a+Ex)[Z(e^{-4V}B\xi t+\cdot\cdot\cdot)
-4g_{4}^{2}W(\xi a't+\delta A'_x)],
\eea
where "$\cdot\cdot\cdot$" represents terms that do not depend on the $t$ coordinate,
resulting in 
\bea
\partial_t(\sqrt{-g}G^{tx})&=&0\cr
\partial_x(\sqrt{-g}G^{xy})&=&\frac{E}{2}(Ze^{-2V}B-4g_{4}^{2}Wa')\cr
\partial_t(\sqrt{-g}G^{ty})&=&-\frac{1}{2}Ze^{-2V}B\xi a+\frac{1}{2}Ze^{-2V}B(-E+\xi a)\nonumber \\
&&-\frac{1}{2}(2a+Ex)(Ze^{-2V}B\xi-4g_{4}^{2}Wa'\xi).
\eea
Note that we consider always the case when $a(x)$ dominates over $Ex$.

Finally, one obtains
\bea
\partial_r(\sqrt{-g}G^{rx})&=&0,\nonumber\\
\partial_r(\sqrt{-g}G^{ry})&=&e^{2V}H^{xy}(E-2\xi a(r))\cr
&=&-(e^{-2V}Z\phi)B-4g_4^2W(\phi)a')(E-2\xi a(r))\;,\label{modifcons}
\eea
where we have used $F^{xy}=Ze^{-4V}B, \sqrt{-g}\tilde F^{xy}=-a'$ and $\sqrt{-g}=e^{2V}$, which can be easily 
calculated.
This in turn is consistent with a particular example of the more general formula presented in 
\cite{Donos:2017mhp}, 
\be
\d_r(\sqrt{-g}G^{ir})=\d_j(\sqrt{-g}G^{ji})+2\sqrt{-g}G^{ij}\xi_j+\sqrt{-g}H^{ij}E_j\;,
\ee
upon specializing to $\xi_i=\xi \delta_{ix},E_i=E\delta_{ix}$ and using $G^{yx}=-aH^{yx}$.

Since there is an extra term in the conservation law (\ref{modifcons}), like in the case of the electric current, we can 
add an extra term to the heat current, obtaining the fluxes (compare with (\ref{heattot}))
\bea
{\cal Q}^x&=&U^2\left(\frac{e^{2V}\delta h_{tx}}{U}\right)^{\prime}-a(r)\sqrt{-g}H^{rx},\nonumber\\ 
{\cal Q}^y&=&U^2\left(\frac{e^{2V}\delta h_{ty}}{U}\right)^{\prime}-a(r)\sqrt{-g}H^{ry}-M(r)E-2M_Q(r)\xi,\label{Hcurr}
\eea
where $M(r)$ and $M_Q(r)$ are given by (\ref{heatmag}) and (\ref{magdensity}), only integrated until $r$ instead of 
infinity. Note that their integrands match the right hand side of the non-conservation in (\ref{modifcons}), so 
by derivating with respect to $r$ we obtain the needed extra term to cancel the non-conservation, so that 
\be
\d_r {\cal Q}^i=0\;,
\ee
as wanted. But by construction $M(r)$ and $M_Q(r)$ (which are integrated from the horizon to $r$) vanish at the horizon. 
Moreover, at the boundary $r\rightarrow\infty$, $M(r)\rightarrow M, M_Q(r)\rightarrow M_Q$, so the extra term are the 
magnetization currents, and subtracting them we obtain the pure transport currents, 
\be
Q^i=Q^{{\rm (tot)}i}-ME-2M_Q\xi={\cal Q}^i(r\rightarrow \infty)={\cal Q}^i(r_H)\;
\ee
so that at the horizon we calculate the transport currents. 

At the horizon, not only $M(r_H)=M_Q(r_H)=0$, but also $a(r_H)=0$ and $U(r_H)=0$ (but 
$U'(r_H)\neq 0$) by the boundary (regularity) condition there, 
which means that finally we obtain 
\be
Q^i=-\left.U' e^{2V}\delta H_{ti}\right|_{r=r_H}.
\ee
This is the same formula as in the case without topological term, in \cite{Blake:2015ina}. The graviton perturbations in the 
presence of $E,B,\xi$ sources was already calculated in (\ref{grav}), so substituting them in the above, and 
comparing with the general formula
\be
Q^i=T\a_{ij}E_j-\kappa_{ij}\nabla_jT\;,
\ee
with $\nabla_i T=\xi \delta_{ix}T, E_i=E\delta_{ix}$ and $U'(r_H)=4\pi T$, we thus extract the coefficients of $TE$ and 
$\xi T$ as 
\bea
\alpha_{xx}&=&\frac{s\rho e^{2V}k^2\Phi}{B^2\rho^2+(B^2Z+e^{2V}k^2\Phi)^2},\nonumber\\ 
\alpha_{xy}&=& sB\frac{(B^2 Z^2+\rho e^{2V}k^2\Phi+\rho^2)}{B^2\rho^2+(B^2Z+e^{2V}k^2\Phi)^2}\cr
\kappa_{xx}&=& \frac{s^2T(B^2Z+e^{2V}k^2\Phi)}{B^2\rho^2+(B^2Z+e^{2V}k^2\Phi)^2}\cr
\kappa_{xy}&=& \frac{s^2T\rho B}{B^2\rho^2+(B^2Z+e^{2V}k^2\Phi)^2}.\label{alphakappa}
\eea
The thermoelectric coefficients agree with the results obtained from the electric current in 
(\ref{etconddxy}), as they should, by general transport theory. We have no new contributions from the topological 
term with $W(\phi)$.

\subsection{S-duality}

The general conductivity formulas (\ref{elconddxy},\ref{etconddxy},\ref{alphakappa}) 
contain explicitly a 
nonzero electric charge $\rho$, and magnetic field $B$, but no nonzero magnetic charge or electric field, so as they 
are, they do not exhibit manifest S-duality (Maxwell duality in a more general setting). However, we can consider 
$\rho=0,B=0$ in them, and obtain 
\bea
\sigma_{xx}&=&Z(r_H)\cr
\sigma_{xy}&=&-4W(r_H)\cr
\a_{xx}&=&0=\a_{xy}=\frac{\kappa_{xy}}{T}\cr
\frac{\kappa_{xx}}{T}&=&\frac{s^2}{e^{2V(r_H)}k^2\Phi(r_H)}.
\eea

We see that the isotropic thermal conductivity $\kappa_{xx}$ is singular for $\Phi(r_H)\rightarrow 0$, but we keep
it finite. In any case, the $\a^{ij}$ and $\kappa^{ij}$ coefficients are invariant under changes of the eletric/magnetic 
variables (S-duality). The other formulas are consistent with previous results at $\rho=B=0$, where we 
know the effect of S-duality \cite{Alejo:2019hnb}. 

Indeed, we can explicitly check that our action (\ref{action}) is invariant under the transformation 
\bea
F_{\mu\nu}&\rightarrow & Z(\phi)\tilde F_{\mu\nu}-\bar W(\phi)F_{\mu\nu}\equiv Z(\phi)\frac{1}{2}\epsilon_{\mu\nu
\rho\sigma}F^{\rho\sigma}-\frac{W(\phi)}{4}\cr
Z(\phi)&\rightarrow & -\frac{Z(\phi)}{Z(\phi)^2+\bar W(\phi)^2}\cr
\bar W(\phi)&\rightarrow& \frac{\bar W(\phi)}{Z(\phi)^2+\bar W(\phi)^2}\;,
\eea
where we have defined $\bar W(\phi)\equiv W(\phi)/4$. 

It was shown in \cite{Alejo:2019hnb} that this transformation comes from a simple duality transformation on the 
action (going to a master action and then writing a dual action in terms of a previously auxiliary field). Moreover, 
as we can see, since $\sigma_{xx}=Z(r_H)$ and $\sigma_{xy}=-\bar W(r_H)$, this transformation becomes 
\bea
\sigma'_{xx}&=&\frac{\sigma_{xx}}{\sigma_{xx}^2+\sigma_{xy}^2}\cr
\sigma'_{xy}&=& -\frac{\sigma_{xy}}{\sigma_{xx}^2+\sigma_{xy}^2}\;,
\eea
or, by defining the usual complex conductivity $\sigma\equiv \sigma_{xy}+i\sigma_{xx}$, simply the usual S-duality formula
acting on complex objects,
\be
\sigma'=-\frac{1}{\sigma}.
\ee

This is indeed the effect of particle-vortex duality (standing in for S-duality in 2+1 dimensions) in the dual field theory, 
as seen for instance in \cite{Burgess:2000kj,Murugan:2014sfa}.

\section{Transport  via entropy function and S-duality}

We next consider an alternative treatment of transport, relevant for {\em extremal} black holes (unlike the nonextremal 
case in the previous section) using the entropy function formalism, and 
generalize the work in \cite{Erdmenger:2015qqa,Erdmenger:2016wyp} to the case with a topological term.

The entropy function formalism was developed by Sen \cite{Sen:2005wa,Sen:2007qy}, having in mind the application 
to the attractor mechanism \cite{Ferrara:1995ih,Ferrara:1996dd}. Within the context of transport, the first application 
was in  \cite{Erdmenger:2016wyp}, whose logic we will follow here.

\subsection{Entropy function formalism}

The entropy function formalism calculates the entropy and other quantities at the horizon of an {\em extremal } 
black hole by the extremization of a function called the entropy function. Since as we saw in the previous section 
often transport properties are determined at the horizon of a black hole in a gravity dual, this formalism will allow us 
to do the calculations easily. 

The specific case we are interested in is the case of an extremal dyonic black hole in four dimensions, which is 
known to have a near-horizon geometry of the type $AdS_2\times S^2$, or $AdS_2\times\mathbb{R}^2$, in the 
case of a planar horizon. The near-horizon metric in this latter (planar) case is 
\be 
ds^{2}=v\left(-r^{2}dt^{2}+\frac{dr^{2}}{r^{2}}\right)+wd\vec{x}^{2}, \label{one}  
\ee
where $v$ is the $AdS_{2}$ radius, $w$ is the $\R^{2}$ radius. The Ricci scalar for this metric is 
\be 
R=-\frac{2}{v}. \label{curvature} 
\ee

The attractor mechanism \cite{Ferrara:1995ih,Ferrara:1996dd} means that the values for the fields at the horizon are 
independent on the values at infinity, depend only on the electric and magnetic charges of the black hole,
and can be found from the extremization of the entropy function. 
For an application in the AdS/CFT correspondence, see \cite{Astefanesei:2007vh}.
The constant values taken by the scalar and vector fields at the horizon are denoted by
\be 
\phi_{s}=u_{s}, \,\,\, F^{(A)}_{rt}=e_{A}, \,\,\, F_{\theta\phi}^{(A)}=B_{A}, \label{fields}
\ee
where ${e_{A}}$ and ${B_{A}}$ are related to the electric and 
magnetic charges respectively.

We define the function $f(u_{s}, v, w, e_{A}, p_{A})$ as the Lagrangian density $\sqrt{- \det g}{\mathcal{L}}$ evaluated for the near-horizon geometry 
(\ref{one}) and integrated over the coordinates of the planar horizon \cite{Sen:2005wa, Sen:2007qy}, 
\be 
f(u_{s}, v_{i}, e_{A}, p_{A})=\int dx dy \sqrt{- \det g}{\mathcal{L}}. 
\ee
Then the entropy function is
\be 
{\mathcal{E}}(\vec{u},\vec{v},\vec{e},\vec{q},\vec{p})\equiv 2\pi[e_{A}Q^{A}-f(\vec{u},\vec{v},\vec{e},\vec{p})]. \label{entrfct}
\ee
Its equations of motion, 
\be 
\frac{\partial {\mathcal{E}}}{\partial u_{s}}=0, \,\,\, \frac{\partial {\mathcal{E}}}{\partial v}=0,
 \,\,\, \frac{\partial {\mathcal{E}}}{\partial w}=0, \,\,\, \frac{\partial {\mathcal{E}}}{\partial e_{A}}=0\;, \label{attracteqs}
\ee
are called attractor equations, and fix the horizon data $(u_s,v,w,e_A)$ as a function of the electric and magnetic 
charges of the black hole, $Q_A,p_A$, thus defining the attractor solution. 

At the extremum (for the true values of the horizon data at the horizon), the entropy function equals the entropy 
of the black hole,
\be 
S_{BH}={\mathcal{E}}(\vec{u},\vec{v},\vec{e},\vec{q},\vec{p}).
\ee
Note that in the case of the $\mathbb{R}^2$ horizon black hole, as $f$ is an integral over the horizon (which has infinite
volume, or rather area), we must consider the entropy density instead.

\subsection{Electrical and heat conductivities}

We want to apply the entropy function formalism, for extremal black holes in an asymptotically AdS gravity dual, 
in order to calculate the transport coefficients, using the formulas 
(\ref{etconddxy},\ref{elconddxy},\ref{alphakappa}).

However, as we mentioned, these results from last section were for nonextremal black holes. But we can consider the 
particular case of extremal black holes by taking the temperature to zero, $T\rightarrow 0$. Indeed, for an extremal 
black hole we have 
\be 
U(r)\approx U(r_{H})+(r-r_{H})U'(r_{H})+\frac{(r-r_{H})^{2}}{2}U''(r_{H})+\mathcal{O}(r^{3})\;,  \label{taylor}
\ee
where $U'(r_H)=4\pi T=0$. Therefore the near-horizon metric is 
\be 
ds^{2}=-\frac{(r-r_{H})^{2}}{2}U''(r_{H})dt^{2}+\frac{2}{(r-r_{H})^{2}U''(r_{H})}dr^{2}
+e^{2V(r_{H})}(dx^{2}+dy^{2})\;, 
\ee
and by the coordinate redefinition
\be 
r-r_{H}= \tilde{\rho}, \,\,\, t= \frac{2}{U''(r_{H})}\tau, \label{change} 
\ee
we obtain the $AdS_2\times \mathbb{R}^2$ metric
\be 
ds^{2}=\frac{2}{U''(r_{H})}\left(-\tilde{\rho}^{2}d\tau^{2}+\frac{d\tilde{\rho}^{2}}{\tilde{\rho}^{2}}\right)
+e^{2V(r_{H})}(dx^{2}+dy^{2}),  \label{adsmetr}
\ee
where therefore
\be
v=\frac{2}{U''(r_H)}\;,\;\;\;
w=e^{2V(r_H)}.
\ee

We can then apply the formalism from the previous section with $T\rightarrow 0$, and then use the entropy function 
formalism from the previous subsection to calculate the horizon data as a function of the electric and magnetic charges. 

Moreover, from the previous section, the ansatz for the field strength to leading order (in the absence of perturbations)
was 
\be 
F=a'(r)dr\wedge dt+B dx\wedge dy.  
\ee
Changing to the near-horizon coordinates, we obtain 
\be 
F=\frac{2a'(r_{H})}{U''(r_H)}d\tilde{\rho}\wedge d\tau + Bdx\wedge dy.  
\ee
Comparing with the ansatz for the entropy function formalism at the horizon, (\ref{fields}), we also 
obtain 
\be
e=\frac{2a'(r_{H})}{U''(r_H)}=va'(r_{H}).\label{identification}
\ee

In order to use the entropy function formalism, 
we consider $\Phi(\phi)=0$ in (\ref{action}), so that we don't have axions, obtaining 
\be 
S=\int d^{4}x\sqrt{-g}\left[\frac{1}{16\pi G_{N}}\left(R-\frac{1}{2}\partial_{\mu}\phi\partial^{\mu}\phi
-V(\phi)\right)-\frac{Z(\phi)}{4g_{4}^{2}}F_{\mu\nu}F^{\mu\nu} -
W(\phi)F_{\mu\nu}\tilde{F}^{\mu\nu}  \right]. \label{actionEMD}
\ee
Using (\ref{one}), (\ref{curvature}) and (\ref{fields}), we compute the Lagrangian in the near-horizon region, 
\be 
\sqrt{-g}{\mathcal{L}}=\frac{1}{16\pi G_{N}}\left(-2w-wvV(u_{D})\right)
+\frac{Z(u_{D})}{2g_{4}^{2}}\left(\frac{w}{v}e^{2}-\frac{v}{w}B^{2}\right)+4W(u_{D})eB\;, 
\ee
where $u_{D}$ is the value of the dilaton field on the horizon.

The entropy function (\ref{entrfct}) is then
\be 
{\mathcal{E}}=2\pi[e_{A}Q^{A}-{\rm Vol}\mathbb{R`}^{2}\sqrt{-g}{\mathcal{L}} ]. \label{entrfct1d}
\ee

The attractor equations (equations of motion of the entropy function) for our system are then
\be  
\frac{Q}{{\rm Vol}\mathbb{R}^{2}}- \frac{Z(u_{D})}{g_{4}^{2}}\frac{w}{v}e-4W(u_{D})B=0,\label{QAd}
\ee
\be 
\frac{Z(u_{D})}{2g_{4}^{2}}\left(\frac{w}{v^{2}}e^{2}+\frac{B^{2}}{w}\right) 
+\frac{w}{16\pi G_{N}}V(u_{D})=0,  \label{dervd}
\ee
\be 
\frac{2}{16\pi G_{N}}- \frac{Z(u_{D})}{2g_{4}^{2}}\left(\frac{1}{v}e^{2}+\frac{v}{w^{2}}B^{2}\right) 
+\frac{v}{16\pi G_{N}}V(u_{D})=0,\label{derwd} 
\ee
\be 
-\frac{1}{2g_{4}^{2}}\frac{\partial Z(u_{D})}{\partial u_{D}}\left(\frac{w}{v}e^{2}
-\frac{v}{w}B^{2}\right)-4\frac{\partial W(u_{D})}{\partial u_{D}}eB
+\frac{wv}{16\pi G_{N}}\frac{\partial V(u_{D})}{\partial u_{D}}=0. \label{derphid}
\ee
Using (\ref{QAd}) we  eliminate $Q$ from (\ref{entrfct1d}), and obtain
\be 
{\mathcal{E}}=2\pi {\rm Vol}\mathbb{R}^{2}\left[\frac{1}{(16\pi G_{N})}\left(2w+wvV(u_{D})\right)
+\frac{Z(u_{D})}{2g_{4}^{2}}\left(\frac{w}{v}e^{2}+\frac{v}{w}B^{2}\right) \right].  
\ee 
We  combine equations (\ref{dervd}) and (\ref{derwd}) and obtain
\be 
V(u_{D})=-\frac{1}{v},\label{sum} 
\ee
\be 
\frac{Z(u_{D})}{2g_{4}^{2}}\left(\frac{e^{2}}{v^{2}}+\frac{B^{2}}{w^{2}}\right)
=\frac{1}{(16\pi G_{N})}\frac{1}{v}\;,   \label{sub}
\ee
and replacing this in (\ref{entrfct1d}), we obtain the entropy (value of the entropy function on the solution of the 
attractor equations)
\be 
{\mathcal{E}}=\frac{4\pi w {\rm Vol}\mathbb{R}^{2}}{16\pi G_{N}}
=\frac{w {\rm Vol}\mathbb{R}^{2}}{4G_{N}}=\frac{A}{4G_{N}}. \label{bhaw}
\ee
This is the expected Hawking formula for the entropy of the black hole, which shows that the attractor mechanism for the 
entropy function does work in this case as well. 

Moving on to the transport, the electric current is defined in the gravity dual as before, as
\be 
\langle J^{\mu}\rangle =\left. \frac{\delta S_{\text{on-shell}}}{\delta \d_r A_{\mu}}\right|_{\text{boundary}}
=\sqrt{-g}\left(\frac{Z(\phi)}{g_{4}^{2}}F^{\mu\nu}+4W(\phi)\tilde{F}^{\mu\nu}\right). 
\ee
As we saw in the previous section, by subtracting a magnetization term that vanishes at the horizon, we obtain the 
pure transport current (not the total one), and the resulting flux is $r$-independent, so can be calculated at the horizon.  
That means that the charge density $J^0\equiv \rho $ of the dual field theory can be calculated at the horizon, 
obtaining \footnote{Remember that $A_{t}=a(r)$ vanishes at the horizon due to the regularity conditions, 
but  $a'(r)$ does not.}
\be 
\rho = \frac{Z( u_{D})wa'(r_{H})}{g_{4}^{2}} +4W(u_{D})B. \,  \label{eqqrho} 
\ee

Replacing (\ref{eqqrho}) in the attractor equation (\ref{QAd}), with the identification (\ref{identification}), we obtain 
that the charge density of the dual field theory $\rho$ equals the charge density of the gravity dual black hole in the 
entropy function formalism, 
\be
\rho=\tilde{Q}\equiv\frac{Q}{{\rm Vol}\mathbb{R}^{2}}.
\ee

Moreover, the entropy density of the dual field theory equals the entropy density of the black hole, which 
because of (\ref{bhaw}) becomes
\be 
s=\frac{4\pi w}{16\pi G_{N}}. \label{eqqents} 
\ee

Replacing these $\rho,s$, together with $T\rightarrow 0, \Phi(\phi)=0$ in
(\ref{etconddxy},\ref{elconddxy},\ref{alphakappa}), gives the finite results
\bea
\sigma_{xx}&=&0, \cr
\sigma_{xy}&=&\frac{\rho}{B}-4W,\cr
\alpha_{xx}&=&0,\cr
\alpha_{xy}&=&\frac{s}{B},\cr
\frac{\bar{\kappa}_{xx}}{T}&=&\frac{s^{2}Z}{g_{4}^{2}\left(\rho^{2}
+\frac{B^{2}Z^{2}}{g_{4}^{4}}\right)}, \cr
\frac{\bar{\kappa}_{xy}}{T}&=&\frac{\rho}{B}\frac{s^{2}}{\left(\rho^{2}
+\frac{B^{2}Z^{2}}{g_{4}^{4}}\right)}\;, \label{transpcoeff}
\eea
where we wrote $\bar \kappa_{ij}/T$, since this is usually the relevant finite quantity. 

While in the above analysis
we have considered the case of nonzero $B$, let us comment on the case
of zero magnetic field. In order to obtain regular planar black holes with only electric charge 
(and finite chemical potential) for the Einstein-Maxwell-dilaton theory, the dilaton potential 
must be non-zero\footnote{One can write the attractor equations for zero magnetic charge and 
non-zero electric charge and find the solutions only in the case when the dilaton potential is nonzero.}.  
As it was pointed out on page 13 of ref.\cite{Erdmenger:2016wyp}, the regularity of the solutions in 
the extremal limit is guaranteed if the attractor equations admit solutions, since they are also 
solutions to the equations of motion with $AdS_{2}\times\mathbb{R}^{2}$ near-horizon geometry. 
So, Sen's formalism is applicable for the Einstein-Maxwell-dilaton theory with only electric charge if 
there is a non-zero dilaton potential, i.e., zero magnetic field and finite chemical potential
(consider eqs. 3.7 and 3.10 in \cite{Erdmenger:2016wyp} relating $\mu\neq 0, B=0$ with $\rho
=\tilde Q\neq 0$ and $v$ finite, which by our eq. (\ref{sum}) means nonzero potential). 
Notice that the coupling $\gamma$ inside the dilaton potential (for instance $\gamma_m$ in 
(\ref{gammam}) below) must be chosen in order to obtain 
regular solutions, since it might be possible to find limits when the solutions are non-regular. 
As it was also pointed
out in the same reference, Sen's formalism is not applicable for general theories with Lifshitz 
symmetry since these don't have an $AdS_{2}\times\mathbb{R}^{2}$, although it might also 
be possible to obtain finite chemical potential with no magnetic charge in this case.

\subsection{Examples}

Finally, since we have obtained the formulas for the transport coefficients as a function of $\rho/B, s/B$ and $W(u_D)$, 
it remains to solve the attractor equations in specific cases, so as to write explicit formulas for the transport coefficients 
as a function only of the charges and the magnetic field $B$. 

\subsubsection{Constant potential and power law topological term}

We consider first the case that the potential is just a constant negative cosmological constant (giving the AdS vacuum at 
infinity), while the topological term is a power law of the kinetic function $Z(\phi)$, 
\be
V(\phi)=\frac{-6}{L^{2}}\;, \;\;\;W(\phi)=\beta Z^n(\phi).
\ee
We manipulate the attractor equations so that we can write $s,\rho, W(u_D)$ in terms of the charges. 

Equation \eqref{sum} gives $v$, which now is a constant,
\be 
v=\frac{L^{2}}{6}. 
\ee

Equation \eqref{QAd} gives
\be 
\tilde{Q}-\frac{Z}{g^2_{4}}\frac{w}{v}e-4\beta Z^{n}B=0\;,
\ee
which can be solved for $e$ as 
\be 
e=\frac{g^2_{4}}{Z}\frac{v}{w}(\tilde{Q}-4\beta Z^nB).\label{e}
\ee

Using 
\be 
\frac{\partial W}{\partial u_D}=\frac{\partial W}{\partial Z}\frac{\partial Z}{\partial u_D}
=\beta nZ^{n-1}\frac{\partial Z}{\partial u_D}\label{WZ}
\ee
in (\ref{derphid}) and (\ref{derwd}), we obtain 
\bea
\frac{Z}{g^2_{4}}\left(\frac{e^2}{v^2}+\frac{B^2}{w^2} \right)-\frac{1}{16\pi G_N}\frac{2}{v}&=&0, \label{II}\\
\frac{Z}{g^2_{4}}\left(\frac{e^2}{v^2}-\frac{B^2}{w^2} \right)+\frac{8\beta nZ^neB}{wv}&=&0.\label{III}
\eea
Substituting $e$ from (\ref{e}) in the above equations, and eliminating $w$ from the two, as
\be
w^{2}=\frac{v}{\alpha}\left[\frac{Z}{g^2_{4}}B^{2}-4\beta nZ^{n}\tilde{Q}B\frac{g^2_{4}}{Z}
+(4\beta Z^n B)^{2}n\frac{g^2_{4}}{Z}\right]\;,\label{w}
\ee
where $\a\equiv  \frac{1}{16\pi G_N}$,
we obtain the 
polynomial equation for $\tilde Q$,
\be
\tilde{Q}^2-\frac{Z^2}{g^4_4}B^2-8\beta\tilde{Q}B(1-n)Z^n+(4\beta B)^2(1-2n)Z^{2n}=0.\label{Zeq1}
\ee

\begin{itemize}

\item The $n=0$ case.

\end{itemize}

In this case, solving (\ref{Zeq1}) gives 
\be 
\frac{Z}{g^2_4}=\pm\left(\frac{\tilde{Q}}{B}-4\beta\right).
\ee
Substituting back into (\ref{w}) and (\ref{e}), we obtain 
\bea 
w&=&\sqrt{\pm \frac{L^{2}(16\pi G_{N})B}{6}[\tilde{Q}-4\beta B]}\cr
e&=&\pm \sqrt{\pm\frac{L^2}{6(16\pi G_{N})}\frac{B}{(\tilde{Q}-4\beta B)}}. 
\eea

Finally now we can put everything back into (\ref{transpcoeff}) and obtain the nonzero transport coefficients 
as a function of the charges as 
\bea
\sigma_{xy}&=&\frac{\tilde{Q}}{B}-4\beta, \cr
\alpha_{xy}&=&4\pi\sqrt{\pm \frac{L^{2}}{6(16\pi G_{N})}\left(\frac{\tilde{Q}}{B}-4\beta \right)},\cr
\frac{\bar{\kappa}_{xx}}{T}&=&(4\pi)^{2}\frac{L^{2}}{6(16\pi G_{N})}\frac{(\tilde{Q}
-4\beta B)^{2}}{\tilde{Q}^{2}+(\tilde{Q}-4\beta B)^{2}}\cr
\frac{\bar{\kappa}_{xy}}{T}&=&\pm(4\pi)^{2}\frac{L^{2}}{6(16\pi G_{N})}\frac{\tilde{Q}(\tilde{Q}
-4\beta B)}{\tilde{Q}^{2}+(\tilde{Q}-4\beta B)^{2}}. \label{kTxyn0}
\eea

\begin{itemize}

\item The $n=1$ case.

\end{itemize}

In this case, solving (\ref{Zeq1}) gives 
\be 
\frac{Z}{g_4^{2}}=\pm\frac{\tilde{Q}}{B}\frac{1}{\sqrt{1+(4\beta g^2_4)^2}}.
\ee
Substituting back into (\ref{w}) and (\ref{e}), we obtain 
\bea 
w^2&=&\frac{ v}{\alpha}\tilde{Q}B4\beta g^2_4\left[\pm \sqrt{1+\frac{1}{(4\beta g^2_4)^2}}-1  \right]\cr
e&=&\sqrt{v\alpha\tilde{Q}B(\pm \sqrt{1+(4\beta g^2_4)^2}-4\beta g^2_4)}.
\eea
Putting everything back into (\ref{transpcoeff}), we obtain the nonzero transport coefficients 
as a function of the charges as 
\bea
\sigma_{xy}&=&\frac{\tilde{Q}}{B}\left(1\mp\frac{4\beta g_{4}^2}{\sqrt{1+(4\beta g_{4}^2)^2}} \right),\cr
\alpha_{xy}&=&4\pi \sqrt{\frac{L^2}{6(16\pi G_{N})}\frac{\tilde{Q}}{B}\left(\pm\sqrt{(4\beta g_{4}^2)^{2}+1}
-4\beta g_{4}^2 \right)}\cr
\frac{\kappa_{xx}}{T}&=&(4\pi)^2\frac{L^2}{6(16\pi G_{N})}\left(\pm\sqrt{1+(4\beta g_{4}^2)^2}
-4\beta g_{4}^2\right)\frac{\sqrt{1+(4\beta g_{4}^2)^2}}{2+(4\beta g_{4}^2)^2}\cr
\frac{\kappa_{xy}}{T}&=&(4\pi)^2\frac{L^2}{6(16\pi G_{N})}\left(\pm\sqrt{(4\beta g^2)^2+1}
-4\beta g_{4}^{2}\right)\frac{1+(4\beta g^2)^2}{2+(4\beta g^2)^2}.
\eea

\subsubsection{Power law potential and power law topological term}

Next we want to consider the more general case when the potential is polynomial, specifically 
\be
V(u_D)=\sum_m \gamma_m Z^m.\label{gammam}
\ee

Now we still have 
\be
v=-\frac{1}{V(u_D)}\;,\label{vV}
\ee
because of (\ref{sum}), just that the right-hand side is not a constant anymore. 
Further,  \eqref{QAd} is unchanged, so we can still solve for $e$ in the same way, obtaining again (\ref{e}).

However, now from  (\ref{derphid}) and (\ref{derwd}), we obtain 
\bea
\frac{2\alpha}{v}-\frac{\tilde{Z}}{2}\left(\frac{e^2}{v^2}+\frac{B^2}{w^2}\right)
+\alpha \sum \gamma_mZ^m&=&0, \label{n3}\\
-\frac{1}{2{g^{2}}_4}\left(\frac{e^2}{v^2}-\frac{B^2}{w^2}\right)\frac{\partial Z}{\partial u_D}
-4\beta nZ^{n-1}\frac{eB}{wv}\frac{\partial Z}{\partial u_D}+\alpha\sum m\gamma_mZ^{m-1}
\frac{\partial Z}{\partial u_D}&=&0.\label{n4}
\eea

Now, if $\frac{\d Z}{\d u}\neq 0$, substituting $e$ from (\ref{e}) in the above equations, and eliminating 
$w$ from the two, we obtain a new polynomial equation for $\tilde Q$, 
\be
(m-2n+1)Z^{2n}-2(m-n+1)\left(\frac{\tilde{Q}}{4\beta B}\right)Z^n+(m-1)\frac{Z^2}{(4\beta g^2)^2}+(m+1)\left(\frac{\tilde{Q}}{4\beta B}\right)^2=0.\label{PolyVZm}
\ee
Moreover, (\ref{n3}) can be used to solve for $w$, if we substitute in it $e$ from (\ref{e}) and $v$ from (\ref{vV}).

\begin{itemize}

\item The $n=0$ case.

\end{itemize}

In this case, solving (\ref{PolyVZm}) leads to 
\be
\frac{Z}{g^2_4}=\pm \sqrt{-\frac{m+1}{m-1}}\left(\frac{\tilde{Q}}{B}-4\beta\right).
\ee

\begin{itemize}

\item The $n=1$ case.

\end{itemize}

In this case, (\ref{PolyVZm}) becomes
\be
(m-1)\left[1+\frac{1}{(4\b g_4^2)^2}\right]Z^2-2m \frac{\tilde Q}{4\b B}Z+(m+1)\frac{\tilde Q^2}{(4\b B)^2}=0.
\ee
For small perturbations, $4\b g_4^2\gg 1$, its solution behaves like 
\be
Z\sim \frac{\tilde{Q}}{4\beta B}\;,
\ee 
but otherwise the full solution is unenlightening. 

In principle we could proceed as before, and solve for $w$ and replace everything in the transport coefficients, but 
the calculations are difficult (we obtain higher order algebraic equations) and the solutions unenlightening.

\subsection{S-duality}

In this case, we have a different limit of the conductivity formulas with respect to the case at section 3, since now we have 
first $\Phi\rightarrow 0, T\rightarrow 0$, and then nonzero $\rho,B,s$ (the opposite of section 3). As mentioned there, 
we cannot check S-duality explicitly on this background, since we have $\rho\neq 0, B\neq 0$, but $\rho_m=0=E$. 
Moreover (and related) we have black holes with $Q\neq 0, B\neq 0$, but $P=0,E=0$.
We can however take the limit (notice the order of limits though, we first took $\Phi\rightarrow 0$, and then took 
$\rho\rightarrow 0$, unlike in section 3) $\rho\rightarrow 0,s\rightarrow 0$ and obtain 
\be
\sigma_{xx}=0\;,\;\;\;
\sigma_{xy}=-4W(r_H)=-\bar W(r_H)\;,\;\;\;
\a_{xx}=0=\a_{xy}=\kappa_{xy}=\kappa_{xx}.
\ee

Then we obtain a subset of the S-duality of section 3, namely 
\be
\bar W \rightarrow \frac{1}{\bar W}\Rightarrow \sigma_{xy}\rightarrow - \frac{1}{\sigma_{xy}}\;,
\ee
namely what we obtain by restricting to $\sigma_{xx}=0$. 

Notice however that we still have $Z(r_H)\neq0$, and that is due to the order of limits we took (the limits are non-commutative).

\section{Transport from Stokes equations and S-duality}

Starting with \cite{Banks:2015wha}, and developed in \cite{Donos:2015bxe,Donos:2017mhp}, the transport coefficients
$(\sigma,\a,\bar\a,\kappa)_{ij}$ for electric and thermal transport were also obtained from a formalism of perturbations 
of black hole solutions that leads to generalized Stokes equations. In the limit when hydrodynamics is valid, it was shown 
in \cite{Banks:2016krz} that the formalism turns into the fluid/gravity correspondence formalism \cite{Bhattacharyya:2008jc}.

Here we will apply the formulas of \cite{Donos:2017mhp} to some one-dimensional lattices and take a relevant $T\rightarrow
0$ limit, with the goal of, in the next section, make some generalizations for that, and use the entropy function formalism 
for a supergravity-inspired model.

\subsection{Stokes equations from black hole horizons}

We consider the action (\ref{action}) at $\Phi(\phi)=0$, i.e., the Einstein-Maxwell-dilaton action (\ref{actionEMD}), 
which has a topological term for the gauge field. 

We consider electrically charged black holes solutions in 3+1 dimensions, with a metric and gauge field 
\bea
ds^2&=&g_{tt}dt^2+g_{rr}dr^2+g_{ij}dx^idx^j+2g_{tr}dtdr+2g_{ti}dtdx^i+2g_{ri}drdx^i,\cr
A&=&A_tdt+A_rdr+A_idx^i.
\eea

At infinity, the solution should go to $AdS_4$ with sources, so 
\bea
ds^2&\rightarrow& r^{-2}dr^2+r^2[g^{(\infty)}_{tt}dt^2+g^{(\infty)}_{ij}dx^idx^j+2g^{(\infty)}_{ti}dtdx^i],\cr
A&\rightarrow& A^{(\infty)}_tdt+A^{(\infty)}_idx^i,\cr
\phi&\rightarrow& r^{\Delta-3}\phi^{(\infty)}\;,
\eea
where $A_t^{(\infty)}=\mu(x)$ is the spatially-dependent chemical potential (source for particle number in the 
dual CFT), $g_{tt}^{(\infty)}=\tilde G(x)$ and 
$g_{ij}^{(\infty)}=\tilde g_{ij}(x)$ define the source for the energy-momentum tensor of the dual CFT, and 
$\phi^{(\infty)}=\tilde\phi(x)$ is a source for the dual scalar operator in the CFT.

The solution should have a horizon at $r=r_H$, and near it, we expect the expansion 
\bea
g_{tt}(r,x)&=&-U(r)(G^{(0)}(x)+...)\cr
g_{rr}(r,x)&=&U^{-1}(r)(G^{(0)}(x)+...)\cr
g_{ti}(r,x)&=&U(r)(g_{tr}^{(0)}(x)+...)\cr
g_{ti}(r,x)&=&U(r)(G^{(0)}(x)\chi_i^{(0)}(x)+...)\cr
A_t(r,x)&=&U(r)\left(\frac{G^{(0)}(x)}{4\pi T}A_t^{(0)}(x)+....\right)\cr
g_{ij}(r,x)&=&h_{ij}^{(0)}(x)+...\cr
g_{ir}(r,x)&=& g_{ir}^{(0)}(x)+...\cr
A_i(r,x)&=&A_i^{(0)}(x)+...\cr
A_r(r,x)&=&A_r^{(0)}(x)+...\cr
\phi(r,x)&=&\phi^{(0)}(x)+...\;,
\eea
where the dots refer to higher orders in $r-r_H$ and, as before, $U(r)=4\pi T(r-r_H)+...$, which means that the fields 
proportional to $U$ vanish at the horizon. The most relevant 
horizon data  are then $T,h_{ij}^{(0)},A_t^{(0)},\chi_i^{(0)}$ and 
$\phi^{(0)}$.

The metric, gauge field and scalar perturbation that introduces sources for the electric and heat currents is 
\bea
\delta(ds^2)&=&\delta g_{\mu\nu}dx^{\mu}dx^{\nu}+2tg_{tt}\xi_idtdx^i +t(g_{ti}\xi_j+g_{tj}\xi_i)dx^idx^j+2t
g_{tr}\xi_i dr dx^i\cr
\delta A&=&\delta a_{\mu}dx^{\mu}-tE_idx^i+tA_t\xi_idx^i,\phantom{.....} \delta\phi\;,\label{perT}
\eea
where as before we have $E_i(x)dx^i$ electric source and $\xi_i(x)dx^i$ thermal gradient, but are considered periodic, 
and closed as one-forms, $dE=0=d\xi$.

Regularity at the horizon $r_H$ gives the conditions
\bea
\delta g_{tt}&=&U(r)(\delta g^{(0)}_{tt}(x)+{\cal O}(r-r_H)), \phantom{.....} \delta g_{rr}=\frac{1}{U(r)}
(\delta g^{(0)}_{rr}(x)+{\cal O}(r-r_H)), \nonumber\\
\delta g_{ij}&=&\delta g^{(0)}_{ij}(x)+\frac{2\ln (r-r_H)}{4\pi T}g_{t(i}\xi_{j)}+{\cal O}(r-r_H),
\phantom{...............}\delta g_{tr}=\delta g^{(0)}_{tr}(x)+{\cal O}(r-r_H),\nonumber\\
\delta g_{ti}&=&\delta g^{(0)}_{ti}(x)+g_{tt}\xi_i\frac{\ln{(r-r_H)}}{4\pi T}+{\cal O}(r-r_H),\cr
\delta g_{ri}&=&\frac{1}{U}(\delta g^{(0)}_{ri}(x)+\frac{\ln (r-r_H)}{4\pi T}g_{tr}\xi_i+{\cal O}(r-r_H)),\nonumber\\
\delta a_t&=&\delta a_t^{(0)}(x)+{\cal O}(r-r_H)\;,\phantom{................}\delta a_r=U^{-1}(\delta a_t^{(0)}(x)+
{\cal O}(r-r_H)\cr
\delta a_i&=&\frac{\ln (r-r_H)}{4\pi T}(-E_i+A_t\xi_i)+\delta a_i^{(0)}(x)+{\cal O}(r-r_H)\;,\cr
\delta \phi&=&\delta\phi^{(0)}(x)+{\cal O}(r-r_H).
\eea

As we already saw, we can define fluxes that are $r$-independent, by subtracting magnetization terms to the total currents, 
and then at the boundary these are just the transport currents, but they can also be calculated at the horizon, where 
the extra terms vanish:
\bea
{\cal J}^i&=&J^{{\rm (tot)}i}-M_{(b)}^{ij}\xi_j\cr
{\cal Q}^i&=&Q^{{\rm (tot)}i}-M_{(b)}^{ij}E_j-2M_{Q(b)}^{ij}\xi_j\;,
\eea
where $(b)$ means for the background (no fluctuations) and
\be
M^{ij}(r)=\int_{r_H}^r dr \sqrt{-g}H^{ij}\;,\;\;\;
M_Q^{ij}=\int_{r_H}^r dr \sqrt{-g} G^{ij}.
\ee
The equality of the transport and horizon currents, via the radially independent fluxes, is written as
\be
J^i={\cal J}^i=J^i_{(0)}\;,\;\;\; Q^i={\cal Q}^i=Q^i_{(0)}\;,
\ee
where the $(0)$ index signifies horizon value. 

Then  \cite{Donos:2017mhp} obtains Stokes equations for a charged ``fluid'' (is a real fluid only in the hydrodynamics
limit, as we said) for the variables $(v_i,p,w)$, standing in for velocity of the fluid, pressure, and (electric) scalar potential, 
respectively, defined as 
\bea
v_i&\equiv&-\delta g^{(0)}_{ti},\cr
p&\equiv& -\frac{4\pi T}{G^{(0)}}\left(\delta g^{(0)}_{rt}-h^{ij}_{(0)}g^{(0)}_{ir}\delta g^{(0)}_{tj}\right)
-h^{ij}_{(0)}\frac{\partial_i G^{(0)}}{G^{(0)}}\delta g^{(0)}_{tj},\nonumber\\
w&\equiv&\delta a^{(0)}_t.
\eea
Here $h^{ij}_{(0)}$ is the inverse metric for $h^{(0)}_{ij}$.

The resulting (generalized) Stokes equations are 
\bea
&&-2\nabla^j\nabla_{(i}v_{j)}+v^j[\nabla_j\phi^{(0)}\nabla_i\phi^{(0)}-4\pi Td\chi^{(0)}_{ji}]
-F^{(0)}_{ij}\frac{J^i_{(0)}}{\sqrt{h^{(0)}}}\nonumber\\
&=&\frac{\rho_H}{\sqrt{h^{(0)}}}(E_i+\nabla_iw)+4\pi T \xi_i-\nabla_ip.\nonumber\\
\nabla_iv^i&=&0, \phantom{.........} \partial_iJ^i_{(0)}=0,\label{Stokes}
\eea
where the local charge density at the horizon (the horizon data for the zeroth component of the electric current) is 
\be
\rho_H\equiv J^t_{(0)}=\sqrt{h^{(0)}}\left(Z^{(0)}A_t^{(0)}-\frac{1}{2}W^{(0)}\epsilon^{ij}F^{(0)}_{ij}\right)\;,
\ee
we can define a magnetic field at the horizon by 
\be
B_H\equiv \sqrt{h^{(0)}}\frac{1}{2}\epsilon^{ij}F_{ij}^{(0)}\;,
\ee
$W^{(0)}=W(\phi^{(0)})$ is the horizon data for the coefficient of the topological term, 
and the electric and heat currents at the horizon are 
\bea
J^i_{(0)}&=& \rho_H v^i +\sqrt{h^{(0)}}\left(Z^{(0)}h^{ij}_{(0)}-W^{(0)}\epsilon^{ij}\right)\left(E_j+\nabla_j w
+F_{jk}^{(0)}v^j\right)\cr
Q^i_{(0)}&=&4\pi T \sqrt{h^{(0)}}v^i.\label{JQ}
\eea

For a particular case, one can next calculate these currents, and as before, identify the coefficients of $T\xi$ and $TE$
as the transport coefficients.

\subsection{Results for one-dimensional lattices}

Here we mostly follow \cite{Donos:2017mhp}.

The relevant case we are interested in is of one-dimensional lattices, where the only nontrivial dependence is on a 
single coordinate $x$, and the fields are independent of the others. Then, in particular for the spatial metric in boundary 
directions at the horizon (horizon data) we consider 
\be
h^{(0)}_{ij}dx^idx^j=g^{(0)}_{ij}dx^idx^j=\gamma(x)dx^2+\lambda(x)dy^2.
\ee
Then one of the Stokes equations, the incompressibilty condition $\nabla_i v^i=0$ becomes (for a single nonvanishing 
component $v^x$, $0=\nabla_x v^x=\frac{1}{\sqrt{-h}}\d_x(\sqrt{-h}v^x)$, and denoting the constant by $v_0$, we 
solve it by
\be
v^x=(\gamma g_{d-1})^{-1/2}v_0.
\ee

Moreover, we consider also
\be
F_{xy}^{(0)}=B_H(x)\;,\;\;\; 
4\pi T \chi_y(x)=\chi(x)\;,\;\;\;\chi_x=0\;,\;\;\;
\phi^{(0)}=\phi^{(0)}(x)\;\,;\;\;\;
A_t^{(0)}=A_t^{(0)}(x)\;,
\ee
and all the horizon data depending on $x$ are periodic with period $L$. We can define also the average over a period, 
$\int \equiv (1/L)\int_0^L dx$, and then the zero modes
\be
B=\int B_H, \phantom{.....} \rho=\int\rho_H, \phantom{.....} s=\int s_H. 
\ee
Note that the entropy density of the horizon is (by the Hawking formula)
\be
s_H=4\pi \sqrt{\gamma \lambda}.
\ee
Moreover, separate the zero modes of $B_H$ and $\rho_H$, and write the remainder as $\d_x$ of something, defining
\be
B_H=B+\d_x \hat A_y\;,\;\;\;
\rho_H=\rho+\d_x C.
\ee

We also define $x$-dependent averages $\int^x$ as the average with $L$ replaced by $x$ in the upper limit of integration. 
Then consider 
\be
w_1(x)=\rho\left(\frac{1}{B}\int^x B_H-\frac{1}{\rho}\int^x \rho_H\right),
\phantom{....} w_2(x)=Ts\left(\frac{1}{B}\int^xB_H-\frac{1}{s}\int^xs_H\right)\;,
\ee
and then construct the periodic functions
\be
u_i=\int^x\frac{\gamma^{1/2}\Sigma_i}{\lambda^{3/2}}-\frac{\int \frac{\gamma^{1/2}\Sigma_i}
{\lambda^{3/2}}}{\int \frac{\gamma^{1/2}}{\lambda^{3/2}}}\int^x\frac{\gamma^{1/2}}{\lambda^{3/2}}\;, \label{uijij}
\ee
where $\Sigma_i$ stands for the set of periodic functions $(\Sigma_1,\Sigma_2,\Sigma_3,\Sigma_4,\Sigma_5)=
(\chi, w_1, w_2, \hat{A}_y, C)$.

Finally, define the
matrix with constant components
\begin{equation}
\mathcal{U}_{ij}=\int \frac{\lambda^{3/2}}{\gamma^{1/2}}\partial_xu_i\partial_xu_j.
\end{equation}

For the transport coefficients, it turns out that one needs to define also the constant 
\be
X=\int \frac{(\partial_x\lambda)^s}{\lambda^{5/2}\gamma^{1/2}}+\int \frac{(\partial_x\phi^{(0)})^2}
{(\gamma\lambda)^{1/2}}+\int \frac{(\rho_H+B_HW^{(0)})^2}{\lambda Z^{(0)}(\lambda\gamma)^{1/2}}
+\int \frac{B^2_HZ^{(0)}}{\lambda(\lambda\gamma)^{1/2}}+\mathcal{U}_{11}.
\ee

Then one solves the Stokes equations for the velocities $v^i$ and currents $J_{(0)}^i$ as a function of the souces
$E_i,\xi_i$, and extracts the transport coefficients.

\subsubsection{Constant $B_H$, $\gamma(x)=\lambda(x)$ and $T\rightarrow 0$ limit}

The case that we will mostly be interested in is of $B_H(x)=B=$constant and $\lambda(x)=\gamma(x)$. 
The last condition can be thought of as using residual diffeomorphism invariance to fix $\lambda=\gamma$. 

Then we obtain first 
\be
u_i=\int^x\frac{\Sigma_i}{\lambda}-\frac{\int \frac{\Sigma_i}{\lambda}}{\int \frac{1}{\lambda}}\int^x\frac{1}{\lambda}\/,
\ee
and then 
\be
\mathcal{U}_{ij}=\int \partial_xu_i\Sigma_j=\int \frac{\Sigma_i\Sigma_j}{\lambda}
-\frac{\int \frac{\Sigma_i}{\lambda}\int \frac{\Sigma_j}{\lambda}}{\int\frac{1}{\lambda}}.
\ee

Next, we have $s_H=4\pi \lambda$, and then 
\bea
w_1(x)&=&\rho x-\int^x\rho_H,\cr
w_2(x)&=&4\pi T\left(x\int \lambda-\int^x\lambda\right)\cr
X&=&\int \frac{(\partial_x\lambda)^2}{\lambda^3}+\int \frac{(\partial_x\phi^{(0)})^2}{\lambda}
+\int Z^{(0)}A^{(0)2}_t+\int \frac{Z^{(0)}B^2}{\lambda^2}+\int \frac{\chi^2}{\lambda}
-\frac{(\int \frac{\chi}{\lambda})^2}{\int \frac{1}{\lambda}}.\label{wX}
\eea

With the above formulas, putting $\gamma=\lambda$ and $B_H(x)=B$ in the more general formulas obtained in 
\cite{Donos:2017mhp}, we find for the electric conductivities
\bea
\sigma^{xx}&=&0\nonumber\\
\sigma^{yy}&=&\mathcal{U}_{22}+\int Z^{(0)}+\int \frac{(\frac{\rho}{B}
+W^{(0)})^2}{Z^{(0)}}\nonumber\\
&&-\frac{1}{X}\left(\mathcal{U}_{12}-\int\left(\frac{\rho}{B}+W^{(0)}\right)A^{(0)}_t
-\int\frac{BZ^{(0)}}{\lambda}\right)^2.\nonumber\\
\sigma^{xy}&=&-\sigma^{yx}=\frac{\rho}{B}\;,
\eea
for the thermoelectric conductivities 
\bea
\alpha^{xx}&=&\bar{\alpha}^{xx}=0,\nonumber\\
\alpha^{yy}&=&\bar{\alpha}^{yy}=\frac{\mathcal{U}_{23}}{T}+\frac{s}{B}\int \frac{(\frac{\rho}{B}
+W^{(0)})}{Z^{(0)}}\nonumber\\
&&-\frac{1}{X}\left(\mathcal{U}_{12}-\int\left(\frac{\rho}{B}+W^{(0)}\right)A^{(0)}_t-\int\frac{BZ^{(0)}}{\lambda}\right)\left(\frac{\mathcal{U}_{13}}{T}-\frac{s}{B}\int A^{(0)}_t\right),\nonumber\\
\alpha^{xy}&=&\bar{\alpha}^{yx}=\frac{s}{B},\nonumber\\
\alpha^{yx}&=&\bar{\alpha}^{xy}=\frac{4\pi}{X}\left(\mathcal{U}_{12}-\int\left(\frac{\rho}{B}+W^{(0)}\right)
A^{(0)}_t-\int\frac{BZ^{(0)}}{\lambda}\right)\;,
\eea
and for the thermal conductivities 
\bea
\kappa^{xx}&=&\frac{16\pi^2T}{X},\nonumber\\
\kappa^{yy}&=&\frac{\mathcal{U}_{33}}{T}+\frac{s^2T}{B^2}\int \frac{1}{Z^{(0)}}
-\frac{T}{X}\left(\frac{\mathcal{U}_{13}}{T}-\frac{s}{B}\int A^{(0)}_t\right)^2,\nonumber\\
\kappa^{xy}&=&\bar{\kappa}^{yx}=-\frac{4\pi T}{X}\left(\frac{\mathcal{U}_{13}}{T}-\frac{s}{B}\int A^{(0)}_t\right).
\eea

Note that in our case we have
\be
\frac{\rho}{B}+W^{(0)}=\frac{\lambda Z^{(0)}}{B}A_t^{(0)}.
\ee

Finally, for application to the extremal case (which will be done in the next section), we want to take the 
limit $T\rightarrow 0$, and also (see previous sections), we need to consider $\chi=0$, which means that ${\cal U}_{1i}=0$.
Also note that, because of (\ref{wX}), $w_2/T$ remains finite as $T\rightarrow 0$, so then so does ${\cal U}_{23}/T$ and 
${\cal U}_{33}/T^2$.

We obtain for the nonzero electric conductivities
\bea
\sigma^{yy}&=&\mathcal{U}_{22}+\int Z^{(0)}+\int \lambda^2Z^{(0)}A^{(0)2}_t
-\frac{1}{X}\left[\int \frac{\lambda Z^{(0)}}{B}\left(A^{(0)2}_t+\frac{B^2}{\lambda^2}\right)\right]^2\cr
\sigma^{xy}&=&\frac{\int \lambda Z^{(0)}A^{(0)}_t-\int W^{(0)}B_H}{B}=\frac{\rho}{B}\;,
\eea
for the nonzero thermoelectric conductivities
\bea
\alpha^{yx}&=&-\frac{4\pi}{X}\int \frac{\lambda Z^{(0)}}{B}(A^{(0)2}_t+\frac{B^2}{\lambda^2})\cr
\a^{xy}&=&\frac{s}{B}\cr
\a^{yy}&=&\frac{\mathcal{U}_{23}}{T}+\frac{s}{B^2}\left(\int\lambda A^{(0)}_t-\frac{1}{X}\int A^{(0)}_t
\int\lambda Z^{(0)}\left(A^{(0)2}_t+\frac{B^2}{\lambda^2}\right) \right)\/,
\eea
and for the nonzero and finite thermal conductivities $\frac{\kappa^{ij}}{T}$,
\bea
\frac{\kappa^{yy}}{T}&=&\frac{\mathcal{U}_{33}}{T^2}
+\frac{s^2}{B^2}\int \frac{1}{Z^{(0)}}-\frac{1}{X}\frac{s^2}{B^2}\left(\int A^{(0)}_t\right)^2 \nonumber\\
\frac{\kappa^{xy}}{T}&=&\frac{4\pi}{X}\frac{s}{B}\int A^{(0)}_t\cr
\frac{\kappa^{xx}}{T}&=&\frac{16\pi^2}{X}.
\eea

Here $X$ is (for $\lambda=e^{-w}$)
\be
X=\int [e^{-w(x)}((\partial_x w)^2+(\partial_x \phi)^2)+Z^{(0)}(A^{(0)2}_t+e^{-2w(x)}B^2)].
\ee

Also, the finite thermal conductivity at zero electric current  (obtained by putting $J^i=0$, and thus relating the electric field 
with the thermal gradient, and substituting it in the heat current) 
$\kappa^{ij}_{J^i=0}=\kappa^{ij}-T\alpha^{il}(\sigma^{-1})_{lm}\alpha^{mj}$, is 
\bea
\frac{\kappa_{J=0}^{xx}}{T}&=&\frac{1}{T}\left({\kappa}^{xx}-T\alpha^{xy}(\sigma^{-1})_{yx}\alpha^{xy}\right)\cr
&=&\frac{(4\pi)^2}{X}\left[1-\left(\int\lambda\right)^2\frac{X}{\rho B}\right]\cr
\frac{\kappa^{xy}_{J=0}}{T}&=&\frac{1}{T}\left({\kappa}^{xy}-T\alpha^{xy}(\sigma^{-1})_{yx}\alpha^{xy}\right)\cr
&=&\frac{(4\pi)^2}{X}\left[\rho\int A^{(0)}_t- X\int\lambda\right]\frac{\int\lambda}{B\rho}.
\eea

\subsection{S-duality}

The generalized Stokes equations are invariant under an S-duality transformation of the horizon 
data \cite{Donos:2015bxe,Donos:2017mhp}. Indeed, consider the transformation 
\bea
B_H&\rightarrow& \rho_H \phantom{........................}\rho_H\rightarrow -B_H\cr
Z^{(0)}&\rightarrow & \frac{Z^{(0)}}{Z_{(0)}^2+W_{(0)}^2}\;,\;\;\;\;\;
W^{(0)}\rightarrow  -\frac{W^{(0)}}{Z_{(0)}^2+W_{(0)}^2}\cr
(E_i+\nabla_i w)&\rightarrow & -\frac{1}{\sqrt{h^{(0)}}}\epsilon_{ij}J^j_{(0)}\;,\;\;\;
J^i_{(0)}\rightarrow  -\sqrt{h^{(0)}}\epsilon^{ij}(E_j+\nabla_j w).\label{stokesdual}
\eea
Then, it is easy to check that the Stokes equations (\ref{Stokes}) are left invariant. The transformation on $(Z^{(0)},W^{(0)})$
is understood as a transformation that must be performed on the right-hand side of the definition of $J^i_{(0)}$ in 
(\ref{JQ}), together with the transformation of the other horizon data, namely $(B_H,\rho_H,(E_i+\nabla_i w))$, and then by 
again replacing $J^i_{(0)}$ from (\ref{JQ}) in the result, to finally obtain the transformation of $J^i_{(0)}$.

Defining the horizon data and its inverse S-dual, 
\be
D_H=(\rho_H,B_H,Z^{(0)},W^{(0)})\rightarrow D_H'=\left(B_H,-\rho_H,\frac{Z^{(0)}}{Z_{(0)}^2+W_{(0)}^2},
-\frac{W^{(0)}}{Z_{(0)}^2+W_{(0)}^2}\right)\;,
\ee
then the action on the electric and thermal conductivities is (here we define $\epsilon^{xy}=+1$)
\bea
\sigma^{ij}(D'_H)&=& -\epsilon^{ik}\sigma^{-1}_{kl}\epsilon^{lj}\cr
\a^{ij}(D'_H)&=&-\epsilon^{ik}\sigma^{-1}_{kl}(D_H)\a^{lj}(D_H)\cr
\bar\a^{ij}(D'_H)&=&-\bar \a^{ik}(D_H)\sigma^{-1}_{kl}(D_H)\epsilon^{lj}\cr
\kappa^{ij}(D'_H)&=&\kappa_{J=0}^{ij}(D_H)\;,\label{conductS}
\eea
where as usual the heat conductivity at zero electrical current is $\kappa_{J=0}^{ij}=\kappa^{ij}-T\bar \a^{ik}
\sigma^{-1}_{kl}\a^{lj}$. 

But if $D_H$ is a solution for horizon data, $D'_H$ is not necessarily also a solution. 
Only if the bulk theory is S-duality invariant, specifically under 
\bea
\phi&\rightarrow & -\phi\cr
Z(\phi)&\rightarrow & \frac{Z(\phi)}{Z^2(\phi)+W^2(\phi)}\cr
W(\phi)&\rightarrow & -\frac{W(\phi)}{Z^2(\phi)+W^2(\phi)}\cr
F_{\mu\nu}&\rightarrow & Z(\phi)\tilde F_{\mu\nu}-W(\phi)F_{\mu\nu}\;,\label{Sbulk}
\eea
which we can check that reduces on the horizon data to (\ref{stokesdual}), is $D'_H$ also a solution, and then 
the transformation (\ref{conductS}) of the transport coefficients is indeed a symemtry of the dual field theory. 

Our action (\ref{action}) certainly falls within that category, since as we saw in section 3, the S-duality (\ref{Sbulk}) is 
an invariance of the action. This matches with the analysis of S-duality in section 3.
We will consider more such bulk theories, inspired from ones arising from supergravity, in the next section.

\section{Supergravity-inspired model and generalizations of transport relations for entropy function formalism}

We now consider, as an example, 
a supergravity-inspired model that contains several scalar fields and a potential for them that is 
polynomial in the field. 

Consider the action for $U(1)^4$ gauge fields $A_\mu^I$ coupled to scalars $X_I$ and gravity,
\bea
S&=&\int d^4x\sqrt{-g}\left[\frac{1}{16\pi G_N}\left(R-\frac{1}{32}\left(3\sum_{I=1}^4 (\partial_ {\mu}\lambda_I)^2
-2\sum_{I<J}\partial_{\mu}\lambda_I\partial^{\mu}\lambda_J\right)-V(X)\right)\right.\cr
&&\left.-\frac{1}{4g_4^2}\sum_{I=1}^4 Z_I(X)(F^I_{\mu\nu})^2-\sum_{I=1}^4 
W_I(X)F^I_{\mu\nu}\tilde{F}^{\mu\nu I}\right]\;,
\eea
where $I=1,2,3,4$ labels the scalars $X_I$, subject to the constraint
\be
X_1X_2X_3X_4=1\;,
\ee
the $\lambda_I$ are redefinitions of $X_I$ via 
\be
\frac{X_I}{\sqrt{8}}=e^{-\frac{\lambda_I}{2}}\;,
\ee
the field strengths of the abelian vectors are as usual $F_{\mu\nu}^I=\d_\mu A_\nu^I-\d_\nu A_\mu^I$, and the 
potential for the scalar fields is 
\be
V(X)=-\frac{g^2_4}{4}\sum_{I<J}\frac{1}{X_IX_J}.
\ee

This is  a generalization of the $U(1)^4$ gauged supergravity model, obtained by dimensional reduction of 11 dimensional 
supergravity on $S^7$ and truncation to the Cartan sector in \cite{Duff:1999gh}, 
and which has been considered in the entropy function formalism 
in \cite{Goulart:2016cuv}. To restrict to that model, we put $W_I=0$ and $Z_I=X_I^2$. The generalization considered here
is consistent with the rest of the paper, having arbitrary $Z(\phi),W(\phi)$, only now generalized to a sum over $I=1,2,3,4$. 
To completely generalize, we would consider an arbitrary  potential $V(X)$, but instead we want to keep the features 
of the supergravity truncation. For the same reason, we also keep the constraint $X_1X_2X_3X_4=1$. Note that 
taking $g\rightarrow 0$ leads to the vanishing of the scalar potential, so that is another situation that can be analyzed.

\subsection{Entropy function formalism and solution in terms of charges}

We follow the same method for the entropy function with the attractor mechanism considered in section 4. The 
near-horizon geometry of an extremal planar black hole solution of this model will again be 
$AdS_2\times \mathbb{R}^2$, using the same general ansatz (\ref{metdyonic}) for the solution as in the rest of the 
paper. Note that because we consider the planar horizon case (with $\mathbb{R}^2$ factor) instead of the spherical 
horizon case (with $S^2$ factor) as in \cite{Goulart:2016cuv}, the entropy function and attractor equations will differ from 
that paper, not only by the topological term, but also by the absence of the $2/v_2$ term coming from the Ricci scalar of 
the horizon factor. In this section we will use the notation of \cite{Goulart:2016cuv} and denote $v$ by $v_1$, and $w$ by 
$v_2$, also since we reserve $w$ for use in one-dimensional lattices.
The horizon data for the abelian vector fields and the scalars is written as
\be
X_I=u_I, \phantom{......}F^I_{rt}=e^I, \phantom{......}F^I_{xy}=p^I,
\ee
and similarly as before, this leads to the entropy function 
\bea
\mathcal{E}&=&2\pi\left\{\sum_{I=1}^4 e_Iq^I- v_1v_2\left[\frac{1}{16\pi G_N}\left(-\frac{2}{v_1}-V(X)\right)
\right.\right.\cr
&&\left.\left.+\sum_{I=1}^4\frac{Z_I}{2g^2}\left(\frac{e_I^2}{v^2_1}-\frac{p^2_I}{v^2_2}\right)+4\sum_{I=1}^4 
\frac{W_Ie_Ip^I}{v_1v_2}\right]\right\}.
\eea

The attractor equations derived from it are 
\bea
\frac{\partial\mathcal{E}_B}{\partial e_I}&=&2\pi\left[q^I-v_1v_2\left(\sum_I \frac{Z_I}{g^2_4}\frac{e^2_I}{v^2_1}\right)
-4\sum_I W_Ip^I\right]=0\cr
-\frac{\partial\mathcal{E}_B}{\partial v_1}&=&2\pi\left[\left(+\frac{1}{2g^2_4}
\sum_I Z_I\left(-\frac{v_2}{v^2_1}e^2_I-\frac{p^2_I}{v_2}\right)-\frac{v_2V}{16\pi G_N}
\right)\right]=0\cr
-\frac{\partial\mathcal{E}_B}{\partial v_2}&=&2\pi\left[\frac{-2}{16\pi G_N}+\frac{1}{2g^2_4}\sum_I 
Z_I  \left(\frac{e^2_I}{v_1}+\frac{v_1p^2_I}{v^2_2}\right)-\frac{v_1V}{16\pi G_N}\right]=0\cr
-\frac{\partial\mathcal{E}_B}{\partial u_I}&=&2\pi\left[ v_1v_2\left(\frac{1}{2g^2_4}\sum_J\frac{\partial Z_J}{\partial u_I}\left(\frac{e_J^2}{v^2_1}-\frac{p^2_J}{v^2_2}\right)-\frac{1}{16\pi G_N}\frac{\partial V}{\partial u_I}\right)
+4\sum_J \frac{\partial W_J}{\partial u_I}e_Jp^J\right]=0.\cr
&&\label{attrsugra}
\eea

The first equation in (\ref{attrsugra}) can be solved for $e_I$ in terms of the charges and other parameters, as 
\be
e_I=g^2\frac{v_1}{v_2}\frac{1}{Z_I}(q_I-4W_Ip^I).
\ee
Substituting this in the second and third equation in (\ref{attrsugra}), and adding and subtracting the result, we obtain 
\bea
&&-\frac{1}{16\pi G_N}\left(\frac{2}{v_1}\right)
+\sum_I\frac{Z_I}{g^2_4}\frac{p^{I2}}{v^2_2}+g^2_4\frac{1}{v^2_2}\sum_I \frac{(q_I-4W_Ip^I)^2}{Z_I}=
0,\label{+}\\
&&\left(2V+\frac{2}{v_1}\right)\frac{1}{16\pi G_N}=0\Rightarrow V(u_I)=-\frac{1}{v_1}.\label{sumdiff}
\eea
These give the possibility to write 2 of the 3 horizon data, $v_1,v_2,V(u_I)$, as a function of the third, and the charges
$(q_I,p^I)$, and $W_I(u)$.

Finally, one should be able to solve the last of the equations in (\ref{attrsugra}), for polynomial $Z_I=\sum_m c_mu_I^m$
and $W_I=\sum_n d_n u_I^n$, to obtain $u_I$ as a function of the same data, reducing to dependence on the 
charges. However, before that, we would have to remember that we have the constraint $X_1X_2X_3X_4=1$, which 
means that
\be
u_1u_2u_3u_4=1\;,
\ee
and the potential depends only on 3 of them (the independent ones), while the fourth is found from the above constraint. 
For instance, if $u_4$ is taken to be dependent, and solved for, we have
\be
V(u_1,u_2,u_3)=-\frac{g_4^2}{4}\left[u_1u_2+u_2u_3+u_3u_1+\frac{1}{u_1u_2}+\frac{1}{u_2u_3}
+\frac{1}{u_3u_1}\right].
\ee

Alternatively, we could consider the same theory {\em without} the constraint, so $V(u_1,u_2,u_3,u_4)$. In that case, 
we would have 
\be
\sum_I u_I \frac{\d V(u_D)}{\d u_I}=-2V(u_D)\;,
\ee
and, as an example, substituting in (\ref{attrsugra}) a pure power law case, $Z_I=u_I^m,W_I=W_0u_I^n$, after some 
manipulations we would obtain
\bea
&&(4p^I)^2(m-2n-2)u^{2n}_I-8{q}_Ip^I(m-n-2)u^n_I-(m+2)\frac{p^2_I}{g^4_4}u^{2m}_I\cr
&&+(m-2){q}^2_I=0.\label{polyI}
\eea
This would allow us to solve for $u_I$, in terms of the charges and either $v_2$, or $V(u_D)$ (obtainable from the 
previous equations, relating $V(u_D),v_1,v_2$). For example, for $m=2,n=2$, we would obtain 
\be
-u^4_I\left(1+\frac{1}{(4g^2_4W_0)^2}\right)+\frac{q_I}{4p_IW_0}u^2_I
=0\Rightarrow u^2_I=\frac{q_I}{p_I}\frac{4W_0g^4_4}{1+(4W_0g^2_4)^2}.
\ee
This can be then substituted into $V(u_D)$, resulting in 
\be
V=-\frac{g^2_4}{4}\sum_{I<J} \frac{1}{u_Iu_J}
=-\frac{g^2_4}{4}\frac{1+(4W_0g^2_4)^2}{4W_0g^4_4}\sum_{I<J} \sqrt{\frac{p_Ip_J}{\tilde{q}_I\tilde{q}_J}}\;,
\ee
and then from the attractor equations (\ref{sumdiff}), the second fixes $v_1$, 
\be
v_1=-\frac{1}{V}\;,
\ee
and replacing in the first we fix $v_2$,
\be
v^2_2=\frac{2}{\alpha g^2_4}\frac{1}{1+(4W_0g^2_4)^2}\frac{\sum^4_I \tilde{q}_Ip_I}
{\sum_{I<J} \sqrt{\frac{p_Ip_J}{\tilde{q}_I\tilde{q}_J}}}\;,
\ee
finally fixing all horizon data in tersm of the charges.
Then the entropy density at the horizon (minimum of the 
entropy function) would be (Hawking fromula)
\be
s=\frac{4\pi v_2}{16\pi G_N}\equiv 4\pi \a v_2=\frac{4\pi}{g_4}\sqrt{\frac{2\alpha}{1+(4W_0g^2_4)^2}}
\sqrt{\frac{\sum^4_I \tilde{q}_Ip_I}{\sum_{I<J} \sqrt{\frac{p_Ip_J}{\tilde{q}_I\tilde{q}_J}}}}.
\ee

\subsection{Transport formulas for this generalization}

To use the transport formulas from the previous section, we need to generalize them to this case. But since the 
only such generalization is the fact that we have several scalars $X_I$ and gauge fields $A_\mu^I$, the only thing 
we need to be careful about is where to put the sums over $I$. 

The horizon data is 
\bea
\rho_{H,I}&\equiv& J^t_{(0)I}=\sqrt{h^{(0)}}\left(Z_I^{(0)}A_t^{I(0)}-\frac{1}{2}W_I^{(0)}\epsilon^{ij}F^{I(0)}_{ij}\right)\cr
B_{H,I}&\equiv& \sqrt{h^{(0)}}\frac{1}{2}\epsilon^{ij}F_{ij}^{I(0)}\cr
J_I^{i(0)}&=& \rho_{H,I} v^i +\sqrt{h^{(0)}}\left(Z_I^{(0)}h^{ij}_{(0)}-W_I^{(0)}\epsilon^{ij}\right)
\left(E_j^I+\nabla_j w^I+F_{jk}^{I(0)}v^j\right)\;,
\eea
and we can define the sums over $I$ (total value)
\be
\rho_H=\sum_I \rho_{H,I}\;,\;\;\;
B_H=\sum_I B_{H,I}\;,\;\;\;
J^{i(0)}=\sum_I J_I^{i(0)}\/.
\ee
and, in the case of one-dimensional lattices that we will be interested in, the averages
\be
B_I=\int B_{H,I}\;,\;\;\;
B=\int B_H\;,\;\;\;
\rho_I=\int \rho_{H,I}\;,\;\;\;
\rho=\int \rho_H.
\ee

Then we have a multiply-charged  (pseudo-)fluid with variables $(v_i,p,w_I)$, standing in for velocity and pressure of the 
fluid and electric scalar potentials defined by 
\bea
v_i&\equiv&-\delta g^{(0)}_{ti},\cr
p&\equiv& -\frac{4\pi T}{G^{(0)}}\left(\delta g^{(0)}_{rt}-h^{ij}_{(0)}g^{(0)}_{ir}\delta g^{(0)}_{tj}\right)
-h^{ij}_{(0)}\frac{\partial_i G^{(0)}}{G^{(0)}}\delta g^{(0)}_{tj},\nonumber\\
w_I&\equiv&\delta a^{I(0)}_t.
\eea

The resulting Stokes equations are 
\bea
&&-2\nabla^j\nabla_{(i}v_{j)}+v^j[\nabla_j\phi^{(0)}\nabla_i\phi^{(0)}-4\pi Td\chi^{(0)}_{ji}]
-\sum_I F^{I(0)}_{ij}\frac{J^{i I}_{(0)}}{\sqrt{h^{(0)}}}\nonumber\\
&=&\sum_I\frac{\rho_{H,I}}{\sqrt{h^{(0)}}}(E_i^I+\nabla_iw_I)+4\pi T \xi_i-\nabla_ip.\nonumber\\
\nabla_iv^i&=&0, \phantom{.........} \partial_iJ^{iI}_{(0)}=0,\label{StokesGen}
\eea

Next, we consider one-dimensional lattices. As we described, the case we are most interested in is of $\chi=0$ and 
$\hat A_y=C=0$, and moreover, since we want to apply to extremal black holes, of $T\rightarrow 0$. That means that 
among the $\Sigma_i$ we consider nonzero only $\Sigma_2=w_1$ and $\Sigma_3=w_2$, which have now to be 
generalized to $\Sigma_{2I}=w_1^I$ and $\Sigma_3=w_2$, defined as 
\be
w_1^I(x)=\rho\left(\frac{1}{B_I}\int^x B_{H,I}-\frac{1}{\rho,I}\int^x \rho_{H,I}\right),
\phantom{....} w_2(x)=Ts\left(\frac{1}{B}\int^xB_H-\frac{1}{s}\int^xs_H\right).
\ee

That means that the nonzero components of the ${\cal U}_{ij}$ matrix are ${\cal U}_{2I2I},{\cal U}_{2I3}, {\cal U}_{33}$. 
Moreover, as before, the finite values as $T\rightarrow 0$ are ${\cal U}_{2I2I},{\cal U}_{2I3}/T, {\cal U}_{33}/T^2$.

We can next follow the steps outlined in Appendix D of \cite{Donos:2015bxe} in order to solve the Stokes equations
for $J^i_{(0)},v^i$ as a function of the sources $E_i,\xi_i$, and find first $v^x=v_0/\sqrt{-h}$ as before, then 
$v^y$ as a linear function of $v_0$ (involving a sum over $I$), then $J_I^{(0)x},J_I^{y(0)}$ as a linear function of $v_0$; 
and finally $v_0$ is obtained as a sum over $I$. 

We can consider $E_I^i=E^i$ (equal electric fields for the all the four gauge fields), and define 
conductivities by $J^i_I=\sigma^{ij}_IE_j+T\a^{ij}_I \xi$, in which case we obtain the  the electric conductivities
\bea
\sigma^{xx}_I&=&0\nonumber\\
\sigma^{yy}_I&=&\mathcal{U}_{2I2I}+\int Z_I^{(0)}+\int \frac{(\frac{\rho_I}{B_I}
+W_I^{(0)})^2}{Z_I^{(0)}}\nonumber\\
&&-\frac{1}{X}\left(\int\left(\frac{\rho_I}{B_I}+W_I^{(0)}\right)^2\frac{B_I}{\lambda Z_I^{(0)}}
+\int\frac{B_IZ_I^{(0)}}{\lambda}\right)\times\cr
&&\times \sum_J\left(\int\left(\frac{\rho_J}{B_J}+W_J^{(0)}\right)^2\frac{B_J}
{\lambda Z_J^{(0)}}+\int\frac{B_JZ_J^{(0)}}{\lambda}\right)\cr
\sigma_I^{xy}&=&-\sigma_I^{yx}=\frac{\rho_I}{B_I}\;,
\eea
for the thermoelectric conductivities 
\bea
\alpha_I^{xx}&=&\bar{\alpha}_I^{xx}=0,\nonumber\\
\alpha_I^{yy}&=&\bar{\alpha}_I^{yy}=\frac{\mathcal{U}_{2I3}}{T}+\frac{s}{B_I}\int \frac{(\frac{\rho_I}{B_I}
+W_I^{(0)})}{Z^{(0)}}\nonumber\\
&&-\frac{1}{X}\left(\int\left(\frac{\rho_I}{B_I}+W_I^{(0)}\right)^2\frac{B_I}{\lambda Z_I^{(0)}}
+\int\frac{B_IZ_I^{(0)}}{\lambda}\right)
\sum_J\left(\frac{s}{B_J}\int \frac{B_J}{\lambda Z_J^{(0)}}\left(\frac{\rho_J}{B_J}+W_J^{(0)}\right)\right),\nonumber\\
\alpha^{xy}_I&=&\bar{\alpha}_I^{yx}=\frac{s}{B_I},\nonumber\\
\alpha_I^{yx}&=&\bar{\alpha}_I^{xy}=-\frac{4\pi}{X}\left(\int\left(\frac{\rho_I}{B_I}+W_I^{(0)}\right)^2\frac{B_I}{\lambda
Z_I^{(0)}}+\int\frac{B_IZ_I^{(0)}}{\lambda}\right)\;,
\eea
and for the thermal conductivities 
\bea
\frac{\kappa^{xx}}{T}&=&\frac{16\pi^2}{X},\nonumber\\
\frac{\kappa^{yy}}{T}&=&\frac{\mathcal{U}_{33}}{T^2}+\sum_I\frac{s^2}{B_I^2}\int \frac{1}{Z_I^{(0)}}
+\frac{1}{X}\left(\sum_I\frac{s}{B_I}\int \left(\frac{\rho_I}{B_I}+W_I^{(0)}\right)\frac{B_I}
{\lambda Z_I^{(0)}}\right)^2,\nonumber\\
\frac{\kappa^{xy}}{T}&=&\frac{\bar{\kappa}^{yx}}{T}=\frac{4\pi }{X}\sum_I\frac{s}{B_I}\int \left(\frac{\rho_I}{B_I}
+W_I^{(0)}\right)\frac{B_I}{\lambda Z_I^{(0)}}.
\eea

If we consider the total conductivities $\sigma^{ij}$ and $\a^{ij}$, we have an additional sum over $I$ in the 
respective formulas. On the other hand, if we consider only a single nonzero $E_I$ (the previous case), all the formulas 
have no sums at all, and only $I$ indices. 

We should note that we have the choice of whether one of the currents $J_I$, or their sum, refers to 
the electric charge current, since in AdS/CMT one takes a phenomenological approach, so {\em any} gauge current in the 
bulk could a priori stand for it, either one of the $U(1)^4$ ones, or the diagonal one (the sum of the currents).

Finally, in order to be able to use the results from the previous subsection, we compare the one-dimensional lattice case 
with the set-up for the extremal black hole with $AdS_2\times \mathbb{R}^2$ horizon. First, since the $(x,y)$ space 
corresponds to $\mathbb{R}^2$, we have that
\be
\lambda=v_2.
\ee
That also implies that $\sqrt{h^{(0)}}=\lambda=v_2$.
Second, we have the constant magnetic field at the horizon
\be
B_I=B_{H,I}=\frac{1}{2}\sqrt{h^{(0)}}\epsilon_{ij}F_{ij}^I=v^2p^I.
\ee
Finally, the electric field is (in the gauge $A_r=0$)
\be
G_{rt}^I=\d_r A_t^I=e^I\Rightarrow A_t^I=e^I(r-r_H)\;,
\ee
to be compared with the general formula (for $G^{(0)}=1$) near the horizon,
\be
A_t^I=(r-r_H)(A_t^{(0)}+...)\Rightarrow A_t^{(0)}=e^I\;,
\ee
which finally gives
\be
\rho_I=\rho_{H,I}=\sqrt{h^{(0)}}Z_I^{(0)}A_t^{I(0)}-W_I^{(0)}B_{H,I}
=v_2\left(Z_I^{(0)}e^I-W_I^{(0)}p^I\right).
\ee
With $v_2,e^I$ written in the previous subsection in terms of the charges $q_I,p^I$, this completes calculating the 
transport coefficients in terms of the charges of the dual black holes.

\section{Conclusions}

In this paper we have considered electric and thermal transport, in the presence of magnetic fields and electric charges
and a topological term with coefficient $W$, 
and the effect of S-duality in such theories. We have also found that we can use the entropy
function formalism and the attractor mechanism to give results for the transport coefficients as a function of the 
charges of the black hole in the gravity dual. 

We have found that the only modification of the transport coefficients from previously found formulas is an extra 
term $-4W(r_H)$ in $\sigma_{xy}$, which however means that S-duality acts on the transport coefficients
consistenly with results at $\rho=B=0$. The entropy function formalism was extended to this case, obtaining, 
in conjunction with the general formulas, explicit formulas depending on the charges of the dual black hole. 
S-duality still acts naturally on the transport coefficients, but an order of limits is important now. 

The formalism of Stokes equations for determination of the transport coefficients, 
especially as it applies to one-dimensional lattices, was also considered, and was applied for the case of extremal 
black holes relevant for the entropy function formalism. S-duality is defined now more generally.
A supergravity-inspired model, obtained by extending the 
$U(1)^4$ Cartan subgroup of ${\cal N}=8,d=4$ gauged supergravity in order to make it consistent with the rest of 
the paper, was also considered. The attractor mechanism, used in conjunction with generalized formulas for transport
from Stokes equations, which we obtained, allowed us to write the transport coefficients of this generalized model in 
terms of the charges of the dual black hole.

\section*{Acknowledgements}

We thank Aristomenis Donos for useful discussions. The work of HN is supported in part by CNPq grant 304006/2016-5 and FAPESP grant 2014/18634-9. HN would also
like to thank the ICTP-SAIFR for their support through FAPESP grant 2016/01343-7.
The work of LA is supported by Capes grant 2017/19046-1. The work of PG is supported by FAPESP grant 2017/19046-1. 



\bibliography{Stransport}
\bibliographystyle{utphys}

\end{document}